\documentclass[manuscript]{aastex}
\usepackage{multicol}
\usepackage{threeparttable}
\def\be{\begin{equation}}
\def\ee{\end{equation}}
\def\bea{\begin{eqnarray}}
\def\eea{\end{eqnarray}}
\def\f{\frac}

\usepackage{graphicx}
\usepackage{amssymb}
\usepackage{multirow}
\shorttitle{ENERGY-DEPENDENT GRB PULSE WIDTH DUE TO THE CURVATURE
EFFECT AND INTRINSIC BAND SPECTRUM} \shortauthors{Peng
et al.}

\begin{document}


%\title{Can spectral evolution result in observed spectral lags of Gamma-ray Burst?}  % due to the evolution of rest frame spectral parameters}
\title{ENERGY-DEPENDENT GRB PULSE WIDTH DUE TO THE CURVATURE
EFFECT AND INTRINSIC BAND SPECTRUM}

\author{Z. Y. Peng \altaffilmark{1,2}, X. H. Zhao \altaffilmark{2,3}, Y. Yin\altaffilmark{4}, Y. Y. Bao\altaffilmark{5}, L. Ma \altaffilmark{1,6}}

\altaffiltext{1}{Department of Physics, Yunnan Normal University,
Kunming 650092, China; pzy@ynao.ac.cn}

%\altaffiltext{${\ast}$}{Corresponding author, astromali@126.com }

\altaffiltext{2}{Key Laboratory for the Structure and Evolution of Celestial Objects, Chinese Academy of Sciences, Kunming 650011, China}

\altaffiltext{3}{National Astronomical Observatories/Yunnan
Observatory, Chinese Academy of Sciences, P. O. Box 110, Kunming
650011, China}

\altaffiltext{4}{Department of Physics, Liupanshui Normal College,
Liupanshui 553004, China}

\altaffiltext{5}{Department of Physics, Yuxi Normal College, Yuxi
653100, China}
\altaffiltext{6}{Author to whom correspondence should be addressed}

\begin{abstract}
Previous studies have found that the width of gamma-ray burst (GRB)
pulse is energy dependent and that it decreases as a power-law
function with increasing photon energy. In this work we have investigated
the relation between the energy dependence of pulse and the so-called
Band spectrum by using a sample including 51 well-separated fast
rise and exponential decay long-duration GRB pulses observed by
BATSE (Burst and Transient Source Experiment on the Compton Gamma
Ray Observatory). We first decompose these pulses into rise, and
decay phases and find the rise widths, and the decay widths also
behavior as a power-law function with photon energy. Then we
investigate statistically the relations between the three power-law
indices of the rise, decay and total width of pulse (denoted as
$\delta_r$, $\delta_d$ and $\delta_w$, respectively) and the three
Band spectral parameters, high-energy index ($\alpha$), low-energy
index ($\beta$) and peak energy ($E_p$). It is found that
(1)$\alpha$ is strongly correlated with $\delta_w$ and $\delta_d$
but seems uncorrelated with $\delta_r$; (2)$\beta$ is weakly
correlated with the three power-law indices and (3)$E_p$ does not show
evident correlations with the three power-law indices. We further
investigate the origin of $\delta_d-\alpha$ and $\delta_w-\alpha$.
We show that the curvature effect and the intrinsic Band
spectrum could naturally lead to the energy dependence of GRB pulse
width and also the $\delta_d-\alpha$ and $\delta_w-\alpha$
correlations. Our results would hold so long as the shell emitting
gamma rays has a curve surface and the intrinsic spectrum is a Band
spectrum or broken power law. The strong $\delta_d-\alpha$
correlation and inapparent correlations between $\delta_r$ and three
Band spectral parameters also suggest that the rise and decay phases
of GRB pulses have different origins.

\end{abstract}

\section{INTRODUCTION}
The origin of the gamma-ray bursts (GRBs) still remains unclear though
it has been found for more than 40 years and many progresses have
been made. Generally GRBs have complex temporal profiles with plenty
of pulses and pulse is the basic element of light curves. The pulses
for most bursts overlap each other, while there also exist
well-separated pulses within some bursts. A lot of statistical
studies about pulses have been made because they can provides us
valuable clues to the radiation mechanism and underlying processes
of GRB. In early statistical analysis, the width of GRB pulses was
found to be energy dependent, i.e., the higher energies, the
narrower widthes (e.g., Link, Epstein \& Priedhorsky 1993). Using
the average autocorrelation function to study the average pulse
width, Fenimore et al. (1995) showed that the average pulse width of
many bursts dependence on energy is well fitted by a power-law
function with the power-law index about -0.4. The result was
confirmed by later studies (e.g., Norris et al. 1996; Norris, Marani
\& Bonnell 2000;  Nemiroff 2000; Crew et al. 2003; Norris et al.
2005). Peng et al. (2006) employed two samples consisting of 82
well-separated pulses to test the relation and found a power-law
anti-correlation between the full pulse width and energy. A
power-law correlation between the pulse width ratio and energy is
also seen in the light curves of the majority of bursts in the two
samples within the energy range of BATSE (Burst and Transient Source
Experiment on the Compton Gamma Ray Observatory). Recently it is
found that this power-law relation can be extended to X-ray bands
(see, Zhang \& Qin 2008, Zhang 2008) as well as X-ray flare
(Chincarini et al. 2010). In addition, Zhang \& Qin (2007) showed
that the pulse peak time, the rise time scale, and the decay time scale on
energy are also power-law functions.

The origin of the dependence of the pulse width on photon energy is
still unclear to date. It has been suggested that the power-law
relation could be attributed to synchrotron cooling (e.g., Kazanas,
Titarchuk \& Hua 1998, Chiang 1998; Dermer 1998; Wang et al. 2000),
which gives a slope of -1/2 between the pulse width and photon energy,
similar to the observed -0.4. However, the synchrotron cooling is
unlikely to be responsible for the relation due to its too short
timescale relative to the pulse width. Under the assumption that the
Doppler effect of the relativistically expanding fireball surface
(or, in some papers, the curvature effect) is important, Qin et al.
(2005) showed that, in most cases, this power-law relationship would
exist in a certain energy range, and, within a similar range, a
power-law relation of an opposite trend between the ratio of the
rising width to the decaying width and energy would be expected for
the same burst. Shen, Song \& Li (2005) also considered the
curvature effect and found the slope of the power-law relation to be
-0.1$\thicksim$-0.2, less than the observed -0.4. These suggest that
the curvature effect is an important factor to form the relation.

The observed spectrum of GRB is usually a broken power-law form. The
smooth connection of two power laws is the so-called Band function,
which can described well the observed spectrum (Band et al. 1993). Three
parameters in the Band function are used to characterize the
spectral shape: the low-energy power-law index $\alpha$, the
high-energy power-law index $\beta$, and the peak of the spectral
energy distribution $E_{p}$. It is found that the distributions of
the power-law slopes of pulse width and photon energy obtained by
Peng et al. (2006) and Zhang \& Qin (2007) have large dispersions
for different bursts, while the GRB spectra also vary dramatically
for two different bursts. Whether the energy dependence of pulse is
connected with the observed spectrum in some way or not?  If it is
true, then what does it imply for the mechanism of energy dependence
of GRB pulse and whether does it supply some useful clues to the
origin of the pulse rise and the decay phase? These motivate our
investigations below. In Section 2, we present the sample
description and pulse modeling. The analysis results are given in
Section 3. We give the possible implications of the statistical
correlations in Section 4. Conclusions and discussion are presented
in the last section.

\section{SAMPLE DESCRIPTION AND PULSE MODELING}

The sample we selected comes from Peng et al. (2010) observed by CGRO/BATSE with durations longer than 2 s, which contains 52
individual FRED pulses with the peak fluxes are greater than 1.0 photon $cm^{-2}$ $s^{-1}$ on a 256-ms time-scale (for further information
about the sample selection and spectral analysis, see Kocevski et al. 2003; Peng et al. 2009a; Peng et al. 2010).

In order to investigate the pulse temporal properties, we must model these pulses with a pulse model. A form proportional to the inverse
of the product of two exponentials, one increasing and one decreasing with time, derived by Norris et al. (2005), is a good
model to describe the GRB pulses. Thus we select the pulse model to model our selected pulses, which can be rewritten as
\begin{equation}
I(t)= A \lambda \exp [-\tau_{1} /(t- t_{s})-(t - t_{s})/\tau_{2} ],
\end{equation}
where $t$ is the time since trigger, $A$ is the pulse amplitude, $t_{s}$ is the pulse start time, $\tau_{1}$ and $\tau_{2}$ are the
characteristics of the pulse rise and the pulse decay, and $\lambda =\exp[2(\tau_{1}/\tau_{2})]^{1/2}$.

Similar to Peng et al. (2006, 2009a) and Hakkila et al. (2008) we use the nonlinear least squares routine MPFIT and develop and apply an
interactive IDL routine to fit all of the pulses, which allows the user to set and adjust the initial pulse parameters manually before
allowing the fitting routine to converge on the best-fitting model via the reduced $\chi^{2}$ minimization. For each burst we require
that the signal should be detectable in at least three channels (in this way, the relationship between the pulse width and energy can be
studied). The fits are examined many times to ensure that they are indeed the best ones.

We demonstrate the results with the fitting to GRB 950624b (BATSE trigger 3648), in which three well-separated pulses were observed in
four BATSE energy channels. The narrow distribution of the fitting $\chi^{2}$ values per degree of freedom (see, the right panel in
Figure 1) indicates that the two-exponential model is sufficient to model the pulse light curves.

The parameter values for all identified pulses are obtained, including the pulse peak intensity ($A$), pulse onset time
($t_{s}$), effective onset time ($t_{eff}$), and peak time ($T_{peak}$), as well as the two fundamental timescales ($\tau_{1}$
and $\tau_{2}$) (see Table 2 in Norris et al. 2005). The effective onset time, $t_{eff}$, is defined as the time when the pulse reaches
0.01 times of the peak intensity. Both onset times are relative to the burst trigger time. According to the fitting parameters we can
obtain the measured pulse temporal properties including the pulse width $w = \bigtriangleup \tau _{1/e} =\tau _{2}(1+2\ln
\lambda)^{1/2} $, the pulse rise width, $\tau_{rise} = \frac{1}{2}w(1-k)$, and the decay width
$\tau_{rise} = \frac{1}{2}w(1+k)$. Fitting the pulse width, the pulse rise width, the pulse decay width (in the logarithm) per channel
as a function of the geometric means of the lower and upper BATSE channel boundaries (using 300$-$1000 keV for channel 4) (Norris et al. 2005) shows the
power-law relations indeed exist and the power-law indices are thus obtained (let $\delta _{w}$, $\delta _{r}$, and $\delta _{d}$
denote the indices of the power-law relations between $w$, $\tau_{rise}$, and $\tau_{decay}$ and energy, respectively).
Therefore, the dependence of the three time quantities on energy can be parameterized by the power-law index. Example plots of the
relations between the temporal properties and energy are illustrated in Figure 2 and the three power-law indices are presented in Table 1.

In order to obtain the consistent time intervals between the light
curves and spectra in the same pulse the time-integrated spectra are
reanalyzed for each pulse of our sample based on the analysis of
Peng et al. (2009a), which provided a detailed data description and
spectral modeling of our sample. Only the so-called Band model (Band
et al. 1993) is used to model the pulse spectra in this work. In the
end there are 51 pulses that are included in our analysis after
removing one bad pulse spectrum with Band model in the certain time
interval. The three spectral parameters for all of the pulses are
also listed in Table 1.

%,  Pulse peak times are given by $t_{p}=t_{eff}+ \sqrt{\tau_{1}\tau_{2}}$.

%\begin{figure*}
%%\vspace{202pt}
%\begin{center}
%
%\resizebox{7cm}{5cm}{\includegraphics{Fig1.PS}}
%%\resizebox{7cm}{5cm}{\includegraphics{plag13ccflag5.ps}}
%\resizebox{7cm}{5cm}{\includegraphics{Fig2.PS}}
%\end{center}
%\caption{Example plots of GRB 950624b pulse fits for BATSE channels (left panel) and the histogram of $\chi^{2}$
%in our sample (right panel).}
%\end{figure*}

\begin{figure}
\centering \resizebox{3in}{!}{\includegraphics{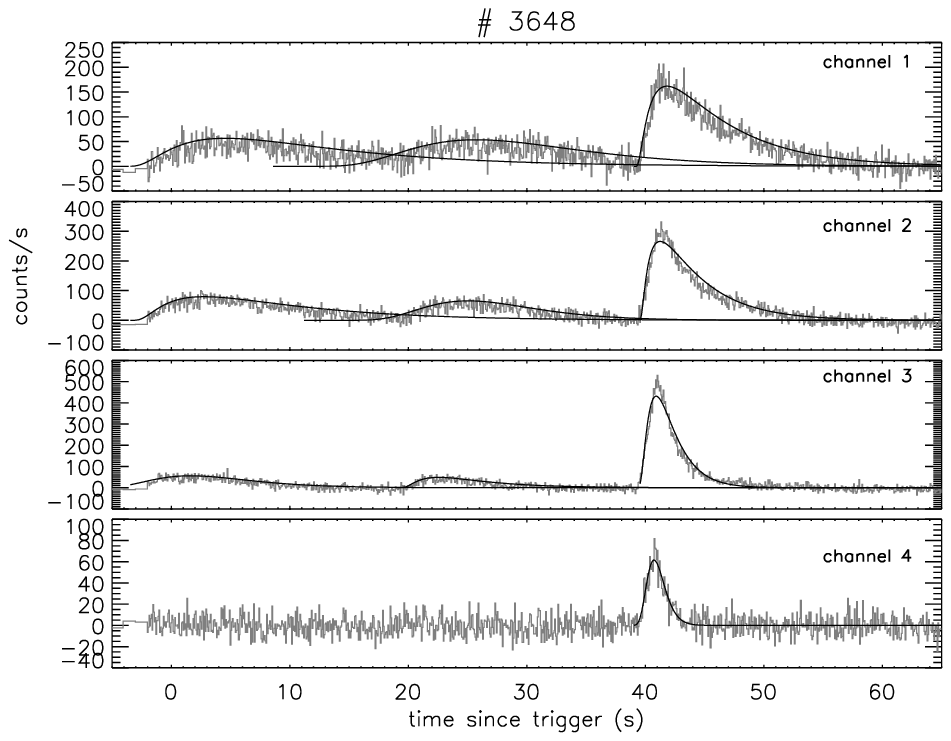}}
\resizebox{3in}{!}{\includegraphics{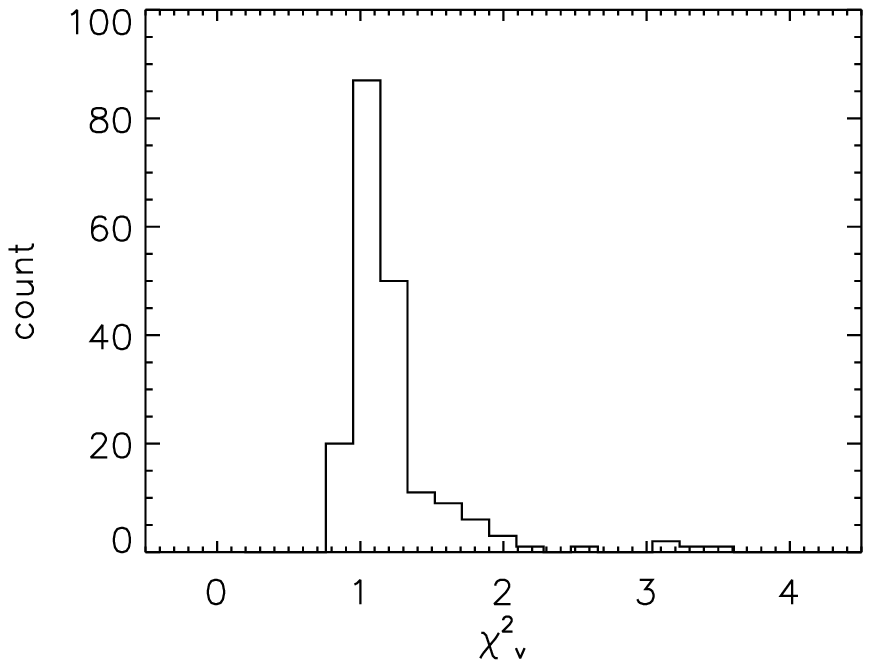}} 
\caption{Example plots of GRB 950624b pulse fits for BATSE channels (left panel) and the histogram of $\chi^{2}$
in our sample (right panel).} \label{}
\end{figure}

\begin{deluxetable}{cccccccc}
\tabletypesize{\scriptsize}  \tablecaption{A List of the Burst Sample
and Various Parameters.
\label{tbl-1}}\tablewidth{0pt} \tablehead{ \colhead{Trigger}
& \colhead{$\delta _{w}$} & \colhead{$\delta _{r}$} & \colhead{$\delta _{d}$}& \colhead{$\alpha$}& \colhead{$\beta$} &\colhead{$E_{p}$ (keV)}}\startdata

%\hline \hline
%\\
563  &     -0.694 $\pm$     0.008  &    -0.566 $\pm$     0.042  &    -0.729 $\pm$     0.019  &    -0.481 $\pm$     0.082  &    -2.669 $\pm$     0.145  &   184.500 $\pm$     7.450 \\
907  &     -0.542 $\pm$     0.016  &    -0.270 $\pm$     0.049  &    -0.647 $\pm$     0.023  &    -0.188 $\pm$     0.065  &    -2.869 $\pm$     0.130  &   188.400 $\pm$     4.740 \\
     914  &     -0.353 $\pm$     0.039  &    -0.408 $\pm$     0.080  &    -0.323 $\pm$     0.020  &    -0.848 $\pm$     0.191  &    -2.475 $\pm$     0.114  &   100.900 $\pm$     6.990 \\
   973$_{-}$1  &     -0.176 $\pm$     0.007  &    -0.170 $\pm$     0.009  &    -0.178 $\pm$     0.006  &    -1.102 $\pm$     0.024  &    -2.084 $\pm$     0.034  &   298.500 $\pm$    12.700 \\
   973$_{-}$2  &     -0.230 $\pm$     0.019  &    -0.176 $\pm$     0.016  &    -0.256 $\pm$     0.024  &    -1.431 $\pm$     0.044  &    -2.264 $\pm$     0.130  &   254.500 $\pm$    25.600 \\
     999  &     -0.250 $\pm$     0.028  &    -0.108 $\pm$     0.047  &    -0.307 $\pm$     0.023  &    -0.971 $\pm$     0.052  &    -2.010 $\pm$     0.067  &   388.800 $\pm$    36.500 \\
 1406  &     -0.379 $\pm$     0.072  &    -0.466 $\pm$     0.082  &    -0.352 $\pm$  0.069  &    -0.616 $\pm$     0.127  &    -2.138 $\pm$     0.033  &   121.700 $\pm$     6.620 \\
    1883  &     -0.207 $\pm$     0.062  &    -0.165 $\pm$     0.047  &    -0.230 $\pm$     0.070  &    -1.294 $\pm$     0.039  &    -3.772 $\pm$     0.850  &   291.000 $\pm$    19.000 \\
    1956  &     -0.163 $\pm$     0.005  &    -0.079 $\pm$     0.008  &    -0.215 $\pm$     0.014  &    -1.132 $\pm$     0.093  &    -2.385 $\pm$     0.117  &   144.300 $\pm$    10.300 \\
    2083  &     -0.406 $\pm$     0.010  &    -0.278 $\pm$     0.045  &    -0.437 $\pm$     0.024  &    -1.326 $\pm$     0.039  &    -3.533 $\pm$     0.191  &    74.380 $\pm$     1.180 \\
    2138  &     -0.408 $\pm$     0.083  &    -0.240 $\pm$     0.010  &    -0.474 $\pm$     0.111  &    -0.315 $\pm$     0.151  &    -2.933 $\pm$     0.408  &   222.900 $\pm$    14.800 \\
    2193  &     -0.582 $\pm$     0.033  &    -0.374 $\pm$     0.025  &    -0.571 $\pm$     0.012  &     0.384 $\pm$     0.064  &    -2.898 $\pm$     0.155  &   274.000 $\pm$     6.480 \\
    2387  &     -0.224 $\pm$     0.009  &    -0.143 $\pm$     0.017  &    -0.261 $\pm$     0.020  &    -0.372 $\pm$     0.048  &    -2.482 $\pm$     0.049  &   158.400 $\pm$     3.200 \\
    2484  &     -0.337 $\pm$     0.010  &    -0.199 $\pm$     0.016  &    -0.407 $\pm$     0.022  &    -0.310 $\pm$     0.123  &    -3.287 $\pm$     0.549  &   178.100 $\pm$     8.220 \\
    2662  &     -0.490 $\pm$     0.019  &    -0.370 $\pm$     0.031  &    -0.544 $\pm$     0.016  &    -0.700 $\pm$     0.129  &    -2.838 $\pm$     0.383  &   178.800 $\pm$    12.500 \\
    2665  &     -0.799 $\pm$     0.128  &    -0.583 $\pm$     0.075  &    -0.890 $\pm$     0.146  &     0.292 $\pm$     0.271  &    -2.856 $\pm$     0.137  &    87.720 $\pm$     3.310 \\
    2700  &     -0.223 $\pm$     0.011  &    -0.534 $\pm$     0.006  &    -0.108 $\pm$     0.012  &    -1.361 $\pm$     0.044  &    -3.005 $\pm$     0.614  &   249.800 $\pm$    16.300 \\
    2880  &     -0.426 $\pm$     0.002  &    -0.269 $\pm$     0.082  &    -0.462 $\pm$     0.021  &    -0.659 $\pm$     0.132  &    -3.114 $\pm$     0.406  &   134.500 $\pm$     6.410 \\
    2919  &     -0.274 $\pm$     0.041  &    -0.222 $\pm$     0.180  &    -0.296 $\pm$     0.025  &    -1.166 $\pm$     0.057  &    -2.253 $\pm$     0.146  &   289.000 $\pm$    28.500 \\
    3003  &     -0.095 $\pm$     0.014  &    -0.058 $\pm$     0.027  &    -0.112 $\pm$     0.009  &    -1.122 $\pm$     0.034  &    -2.053 $\pm$     0.099  &    489.600 $\pm$    47.900 \\
    3143  &     -0.288 $\pm$     0.009  &    -0.147 $\pm$     0.038  &    -0.351 $\pm$     0.030  &    -0.916 $\pm$     0.500  &    -2.032 $\pm$     0.077  &   101.700 $\pm$    26.500 \\
    3256  &     -0.613 $\pm$     0.003  &    -0.653 $\pm$     0.027  &    -0.598 $\pm$     0.006  &    -0.028 $\pm$     0.162  &    -2.939 $\pm$     0.236  &   145.800 $\pm$     6.190 \\
    3257  &     -0.578 $\pm$     0.036  &    -0.449 $\pm$     0.030  &    -0.613 $\pm$     0.039  &    -0.231 $\pm$     0.065  &    -3.040 $\pm$     0.226  &   195.900 $\pm$     5.160 \\
    3415  &     -0.193 $\pm$     0.021  &    -0.041 $\pm$     0.013  &    -0.254 $\pm$     0.034  &    -1.000 $\pm$     0.158  &    -2.093 $\pm$     0.103  &   182.500 $\pm$    27.000 \\
  3648$_{-}$2  &     -0.528 $\pm$     0.012  &    -0.626 $\pm$     0.060  &    -0.479 $\pm$     0.013  &    -0.862 $\pm$     0.333  &    -2.232 $\pm$     0.064  &    98.210 $\pm$    10.000 \\
  3648$_{-}$3  &     -0.528 $\pm$     0.031  &    -0.273 $\pm$     0.027  &    -0.625 $\pm$     0.028  &    -0.748 $\pm$     0.052  &    -2.802 $\pm$     0.155  &   217.200 $\pm$     7.180 \\
    3765  &     -0.231 $\pm$     0.038  &    -0.257 $\pm$     0.165  &    -0.227 $\pm$     0.015  &    -0.872 $\pm$     0.022  &    -2.733 $\pm$     0.101  &   308.800 $\pm$     7.750 \\
    3875  &     -0.425 $\pm$     0.024  &    -0.280 $\pm$     0.142  &    -0.464 $\pm$     0.063  &    -1.245 $\pm$     0.333  &    -2.974 $\pm$     0.219  &    56.380 $\pm$     6.090 \\
    3954  &     -0.117 $\pm$     0.013  &    -0.253 $\pm$     0.022  &    -0.064 $\pm$     0.023  &    -1.153 $\pm$     0.059  &    -1.927 $\pm$     0.049  &   295.500 $\pm$    32.100 \\
    4350  &     -0.136 $\pm$     0.052  &    -0.389 $\pm$     0.159  &    -0.047 $\pm$     0.012  &    -1.467 $\pm$     0.105  &    -2.357 $\pm$     0.420  &   263.500 $\pm$    62.500 \\
    5478  &     -0.427 $\pm$     0.006  &    -0.225 $\pm$     0.117  &    -0.496 $\pm$     0.033  &    -0.287 $\pm$     0.107  &    -2.896 $\pm$     0.265  &   158.100 $\pm$     6.410 \\
    5517  &     -0.288 $\pm$     0.070  &    -0.230 $\pm$     0.026  &    -0.321 $\pm$     0.094  &    -1.156 $\pm$     0.154  &    -2.667 $\pm$     0.000  &   149.700 $\pm$    14.500 \\
    5523  &     -0.269 $\pm$     0.009  &    -0.231 $\pm$     0.018  &    -0.290 $\pm$     0.004  &    -1.266 $\pm$     0.173  &    -2.382 $\pm$     0.188  &   129.200 $\pm$    15.900 \\
    5601  &     -0.335 $\pm$     0.024  &    -0.246 $\pm$     0.041  &    -0.380 $\pm$     0.030  &    -0.725 $\pm$     0.057  &    -2.295 $\pm$     0.076  &   223.200 $\pm$    10.300 \\
    6159  &     -0.155 $\pm$     0.041  &     0.035 $\pm$     0.008  &    -0.237 $\pm$     0.052  &    -0.907 $\pm$     0.446  &    -2.260 $\pm$     0.039  &    73.200 $\pm$     8.610 \\
    6397  &     -0.215 $\pm$     0.008  &    -0.116 $\pm$     0.014  &    -0.259 $\pm$     0.017  &    -0.648 $\pm$     0.042  &    -2.572 $\pm$     0.104  &   192.500 $\pm$     5.440 \\
    6504  &     -0.634 $\pm$     0.037  &    -0.475 $\pm$     0.005  &    -0.687 $\pm$     0.049  &     0.067 $\pm$     0.099  &    -2.729 $\pm$     0.156  &   157.200 $\pm$     4.970 \\
    6621  &     -0.112 $\pm$     0.013  &    -0.074 $\pm$     0.035  &    -0.126 $\pm$     0.009  &    -1.086 $\pm$     0.074  &    -2.424 $\pm$     0.079  &   128.900 $\pm$     5.960 \\
    6625  &     -0.304 $\pm$     0.001  &    -0.253 $\pm$     0.009  &    -0.331 $\pm$     0.006  &    -0.873 $\pm$     0.120  &    -2.841 $\pm$     0.124  &    76.900 $\pm$     2.000 \\
    6657  &     -0.571 $\pm$     0.051  &    -1.017 $\pm$     0.258  &    -0.521 $\pm$     0.058  &    -1.043 $\pm$     0.420  &    -2.334 $\pm$     0.159  &    95.180 $\pm$    13.100 \\
    6930  &     -0.264 $\pm$     0.016  &    -0.171 $\pm$     0.021  &    -0.314 $\pm$     0.012  &    -0.639 $\pm$     0.131  &    -2.294 $\pm$     0.046  &    95.180 $\pm$     4.340 \\
    7293  &     -0.495 $\pm$     0.014  &    -0.281 $\pm$     0.062  &    -0.585 $\pm$     0.039  &     0.313 $\pm$     0.089  &    -2.979 $\pm$     0.120  &   161.300 $\pm$     3.460 \\
    7295  &     -0.410 $\pm$     0.013  &    -0.319 $\pm$     0.020  &    -0.443 $\pm$     0.022  &     0.404 $\pm$     0.094  &    -2.628 $\pm$     0.180  &   373.500 $\pm$    16.500 \\
    7475  &     -0.109 $\pm$     0.010  &    -0.141 $\pm$     0.012  &    -0.092 $\pm$     0.010  &    -1.315 $\pm$     0.066  &    -1.979 $\pm$     0.029  &   149.900 $\pm$    14.300 \\
    7548  &     -0.227 $\pm$     0.001  &    -0.296 $\pm$     0.051  &    -0.195 $\pm$     0.026  &    -0.836 $\pm$     0.104  &    -2.250 $\pm$     0.000  &   176.800 $\pm$    11.100 \\
    7588  &     -0.210 $\pm$     0.024  &    -0.423 $\pm$     0.026  &    -0.134 $\pm$     0.027  &    -0.729 $\pm$     0.208  &    -2.710 $\pm$     0.097  &    71.290 $\pm$     2.100 \\
    7638  &     -0.486 $\pm$     0.012  &    -0.517 $\pm$     0.010  &    -0.480 $\pm$     0.013  &    -1.298 $\pm$     0.389  &    -2.576 $\pm$     0.090  &    54.170 $\pm$     5.300 \\
    7648  &     -0.690 $\pm$     0.197  &    -0.826 $\pm$     0.292  &    -0.647 $\pm$     0.162  &    -0.588 $\pm$     0.100  &    -2.356 $\pm$     0.000  &   201.100 $\pm$    10.300 \\
    7711  &     -0.272 $\pm$     0.031  &    -0.400 $\pm$     0.025  &    -0.219 $\pm$     0.032  &    -1.241 $\pm$     0.051  &    -3.124 $\pm$     0.806  &   211.200 $\pm$    12.800 \\
  8049$_{-}$1  &     -0.208 $\pm$     0.003  &    -0.289 $\pm$     0.034  &    -0.156 $\pm$     0.018  &    -0.391 $\pm$     0.104  &    -3.826 $\pm$     0.550  &   164.300 $\pm$     6.330 \\
  8049$_{-}$2  &     -0.324 $\pm$     0.032  &    -0.430 $\pm$     0.012  &    -0.252 $\pm$     0.063  &    -1.408 $\pm$     0.072  &    -3.096 $\pm$     0.410  &   157.400 $\pm$    12.100 \\
%\hline
\enddata
\end{deluxetable}

\section{ANALYSIS RESULTS}

\subsection{THE RELATIONS AMONG THE POWER-LAW INDICES}
Employing a sample consisting of 24 long-lag pulses Zhang \& Qin (2007) studied the relations between the pulse temporal properties
(width, rise width, decay width and peak time) and energy and found the pulse temporal properties are power-law functions of energy. In
this section, we recheck the relations for the following two reasons. Firstly, the sample we used is much larger than that of
Zhang \& Qin (2007). Secondly, the FRED bursts are temporally and spectrally distinguished from long-lag bursts (Peng et
al. 2010).
%
%\begin{figure*}
% \vspace{202pt} %{plag13windexplot.ps}
% \begin{center}
%\resizebox{7cm}{5cm}{\includegraphics{Fig3.PS}}
%\resizebox{7cm}{5cm}{\includegraphics{Fig4.PS}}
%\resizebox{7cm}{5cm}{\includegraphics{Fig5.PS}}
%\resizebox{7cm}{5cm}{\includegraphics{Fig6.PS}}
%\end{center}
%\caption{Example plots of the relations between the observed pulse temporal properties and energy for our selected samples,
%where the open circle, the diamond, the triangle and the square represent the pulse width, the pulse rise width, the pulse decay
%width, respectively. The power-law relations between them are evident.}
%\end{figure*}

\begin{figure}
\centering 
\resizebox{7cm}{5cm}{\includegraphics{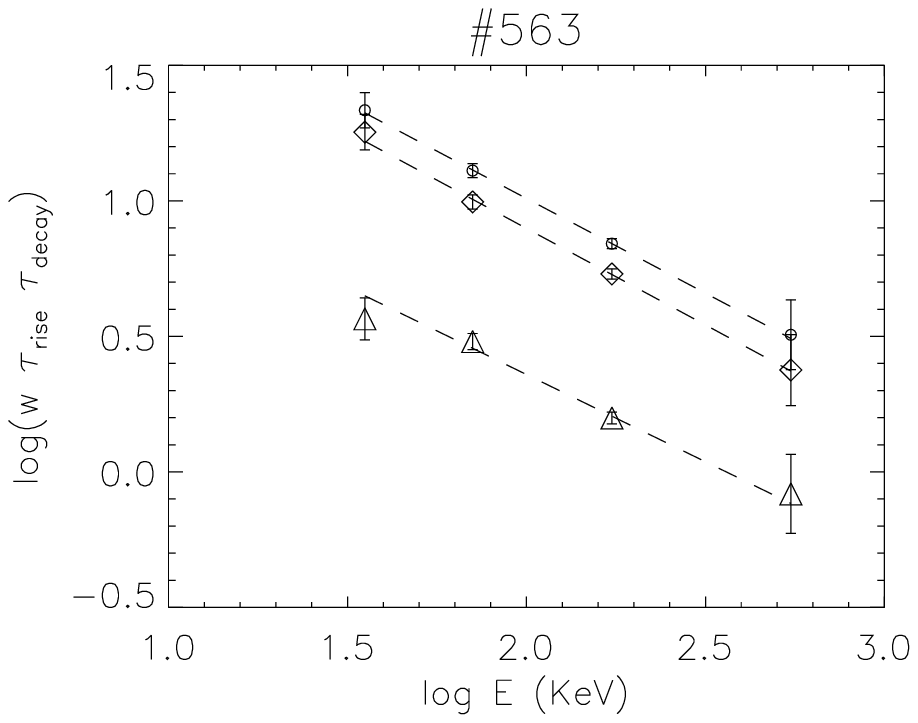}}
\resizebox{7cm}{5cm}{\includegraphics{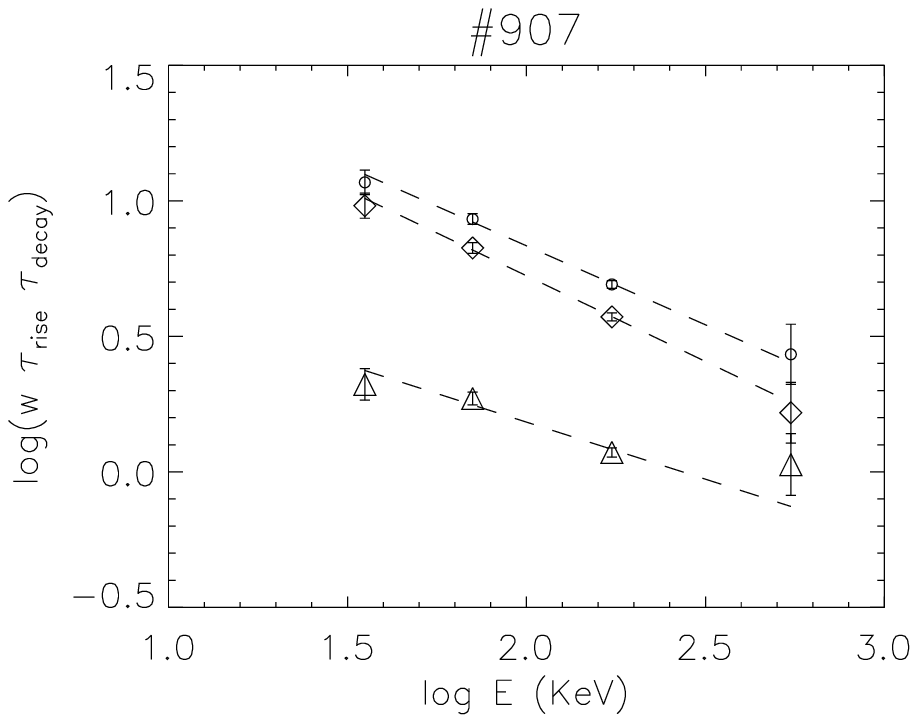}}
\resizebox{7cm}{5cm}{\includegraphics{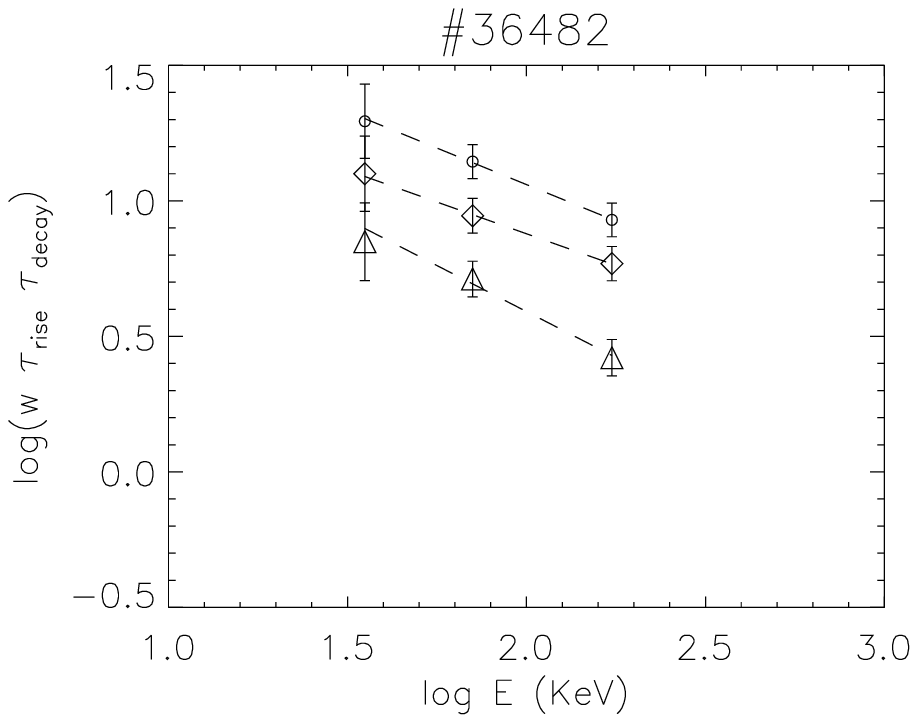}}
\resizebox{7cm}{5cm}{\includegraphics{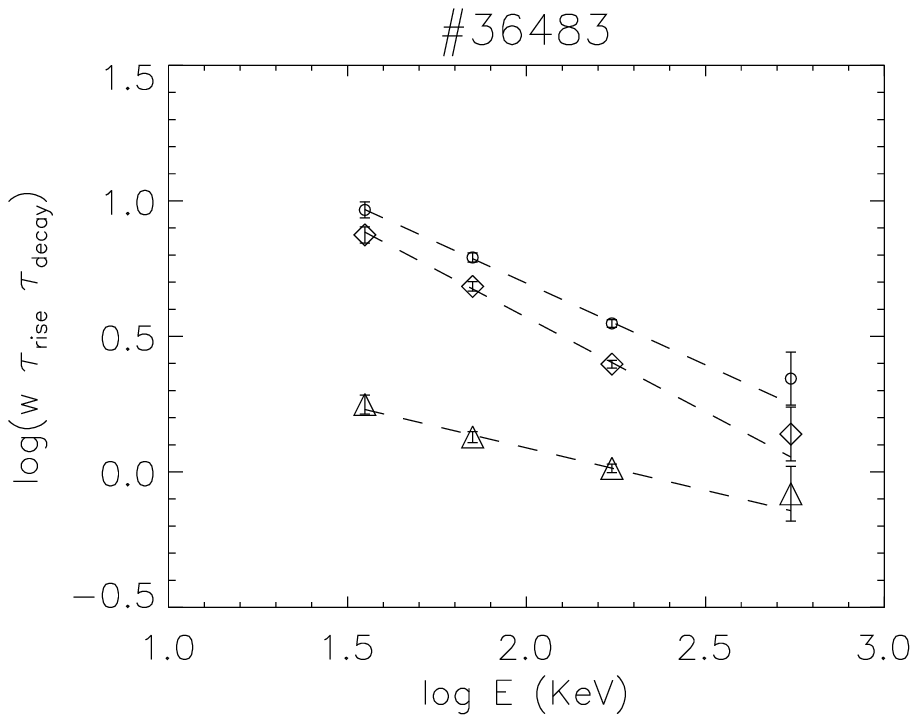}}
\caption{Example plots of the relations between the observed pulse temporal properties and energy for our selected samples,
where the open circle, the diamond, the triangle and the square represent the pulse width, the pulse rise width, the pulse decay
width, respectively. The power-law relations between them are evident.} 
\label{}
\end{figure}

Displayed in Figure 3 are the histograms of the three indices. The distribution parameters are listed in Table 2. From Figure 3 and
Table 2 we find the distributions of these indices share the similar distribution width but they have large dispersions. The large
dispersions imply that the energy dependence of the temporal properties may not be the same for different bursts. In addition, it
is found that most of the power-law indices are much smaller than those of long-lag pulse derived by Zhang \& Qin (2007).

%\begin{figure*}
% \vspace{202pt} %{plag13windexplot.ps}
% \begin{center}
%\resizebox{5.25cm}{5cm}{\includegraphics{Fig7.PS}}
%\resizebox{5.25cm}{5cm}{\includegraphics{Fig8.PS}}
%\resizebox{5.25cm}{5cm}{\includegraphics{Fig9.PS}}
%\end{center}
%\caption{Histograms of the power-law indices $\delta _{w}$ (a), $\delta _{r}$ (b), and $\delta _{d}$ (c) obtained by fitting the
%pulse width, rise width, and decay width and energy with power-law functions, respectively. The dashed lines are the best fits by the
%Gaussian functions.}
%\end{figure*}

\begin{figure}
\centering
\resizebox{5.25cm}{5cm}{\includegraphics{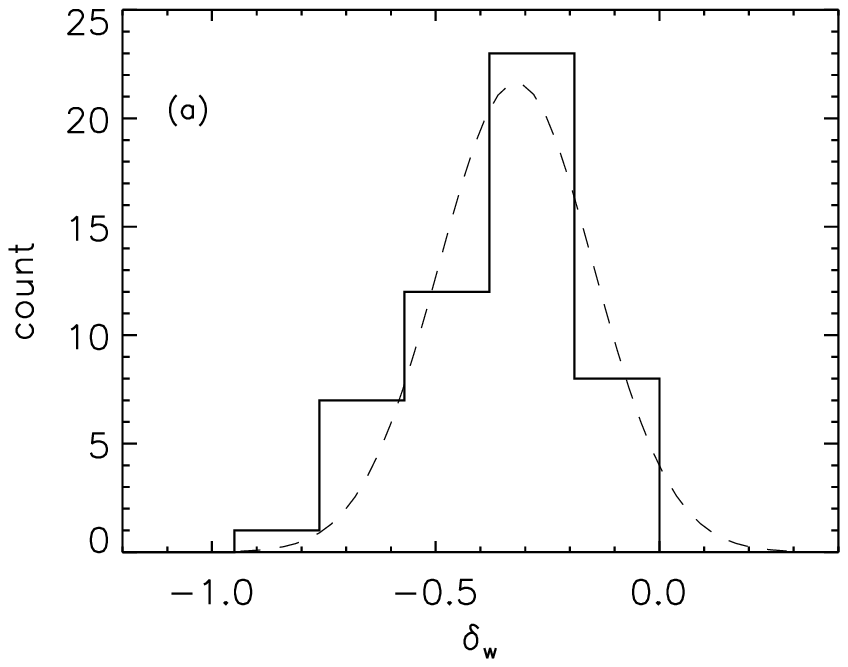}}
\resizebox{5.25cm}{5cm}{\includegraphics{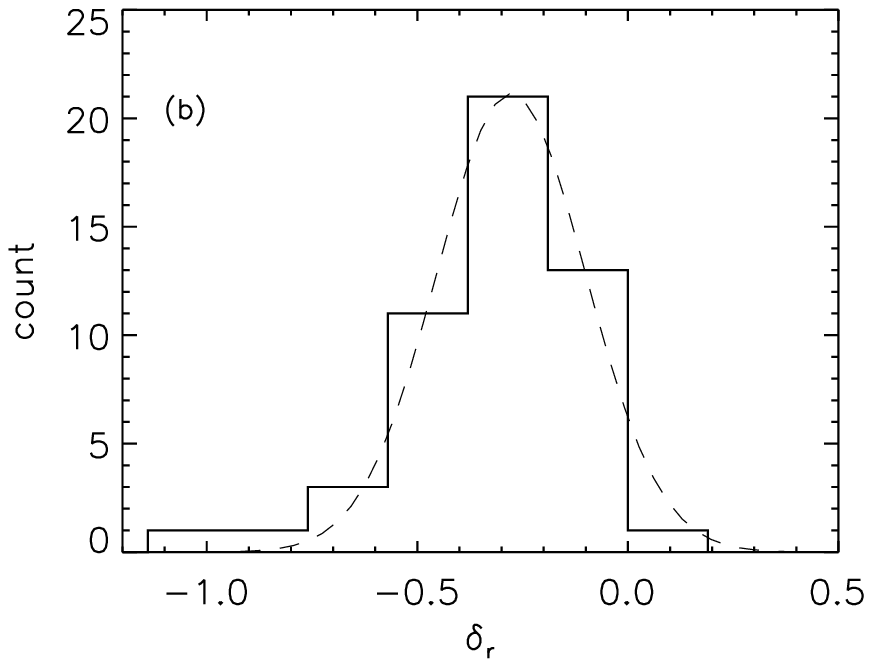}}
\resizebox{5.25cm}{5cm}{\includegraphics{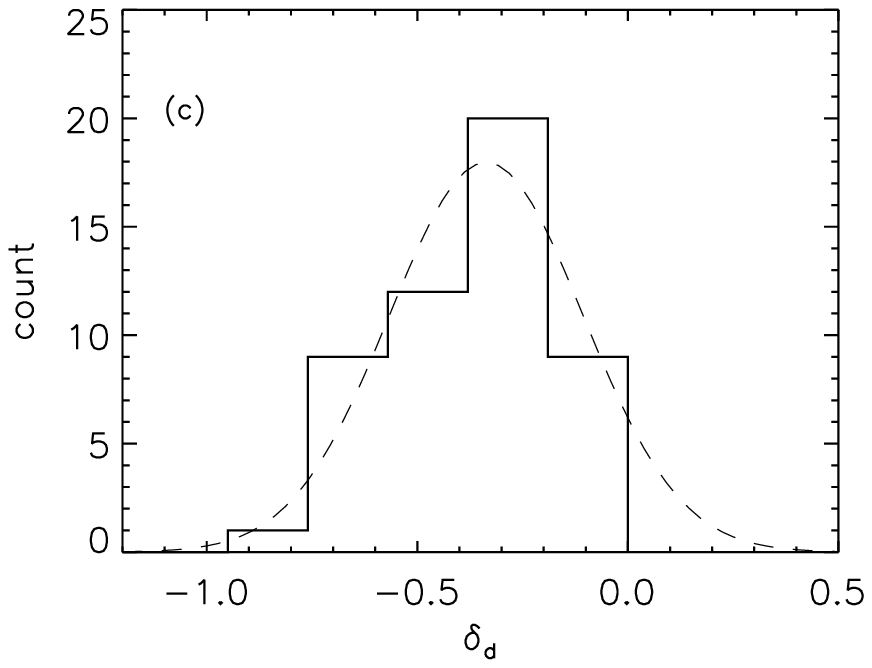}}
\caption{Histograms of the power-law indices $\delta _{w}$ (a), $\delta _{r}$ (b), and $\delta _{d}$ (c) obtained by fitting the
pulse width, rise width, and decay width and energy with power-law functions, respectively. The dashed lines are the best fits by the
Gaussian functions.}
\label{}
\end{figure}

\begin{deluxetable}{cccc}
\tabletypesize{\scriptsize}  \tablecaption{A List of the Distribution Parameters of the Three Power-law Indices.
\label{tbl-2}}\tablewidth{0pt} \tablehead{\colhead{Power-law indices} & \colhead{ Mean} & \colhead{ Median} & \colhead{ $\sigma$
(modeled with a Gaussian profile)}}\startdata
 $\delta _{w}$ & -0.35 & -0.31 &  -0.32 $\pm$ 0.17 \\
 $\delta _{r}$ & -0.31 & -0.27 &  -0.27 $\pm$ 0.17 \\
 $\delta _{d}$ & -0.37 & -0.32 &  -0.33 $\pm$ 0.23 \\
\enddata
\end{deluxetable}

\begin{deluxetable}{ccc}
\tabletypesize{\scriptsize}  \tablecaption{Correlations of the power-law index pairs.
\label{tbl-3}}\tablewidth{0pt}
\tablehead{\colhead{Power-law indices} & \colhead{$R_{S}$} & \colhead{$P_{S}$} }\startdata
$\delta _{w}$- $\delta _{r}$ & 0.69 & $2.63\times 10^{-8}$\\
$\delta _{w}$- $\delta _{d}$ & 0.96 & $2.93\times 10^{-28}$\\
$\delta _{r}$- $\delta _{d}$ & 0.49 &  $2.84\times 10^{-4}$\\
\enddata
\tablecomments{$R_{S}$ and $P_{S}$ denote the Spearman rank-order correlation coefficient and significance, respectively.}
\end{deluxetable}

We also examine the relationships between the power-law indices of the
temporal properties on energy. Figure 4 shows the relationships
between them. The Spearman rank-order correlation analysis of the
three quantities are listed in Table 3. The straight lines are
fitted to the points: (1) $\delta _{r}$= (0.11 $\pm$ 0.01) +
(1.31$\pm$0.03) $\delta _{w}$; (2) $\delta _{d}$= ($-$ 0.07 $\pm$
0.06) + (1.01 $\pm$ 0.02) $\delta _{w}$; (3) $\delta _{r}$= (0.09
$\pm$ 0.01) + (1.25$\pm$0.03) $\delta _{d}$. It is found that the
$\delta_{d}$ is strongly correlated with $\delta _{w}$ and the slope
between them is close to 1. And so the two indices may be viewed as
mutual surrogates. While the other two index pairs are obviously
less correlated, which strongly indicates that the temporal
properties (rise and the decay times) of the pulses do not evolve
independently from each other, instead, their evolution is tightly
coupled.
%The fact that neither the correlation nor the slope between
%$\delta_{w}$ and the two indices $\delta_{r}$, $\delta_{d}$ is the
%same seem to reveal that the rise and decay phases of GRB pulses are
%inherently different.

%which is in good agreement with that of Zhang \& Qin (2007)

%\begin{figure*}
% \vspace{202pt}
% \begin{center}
%\resizebox{5.25cm}{4cm}{\includegraphics{Fig10.PS}}
%\resizebox{5.25cm}{4cm}{\includegraphics{Fig11.PS}}
%\resizebox{5.25cm}{4cm}{\includegraphics{Fig12.PS}}
%\end{center}
%\caption{Relationships of the three power-law indices $\delta _{r}$ vs. $\delta _{w}$ (a), $\delta _{d}$ vs. $\delta _{w}$ (b), $\delta
%_{r}$ vs. $\delta _{d}$ (c). The dashed lines are the regression lines.}
%\end{figure*}

\begin{figure}
\centering
\resizebox{5.25cm}{4cm}{\includegraphics{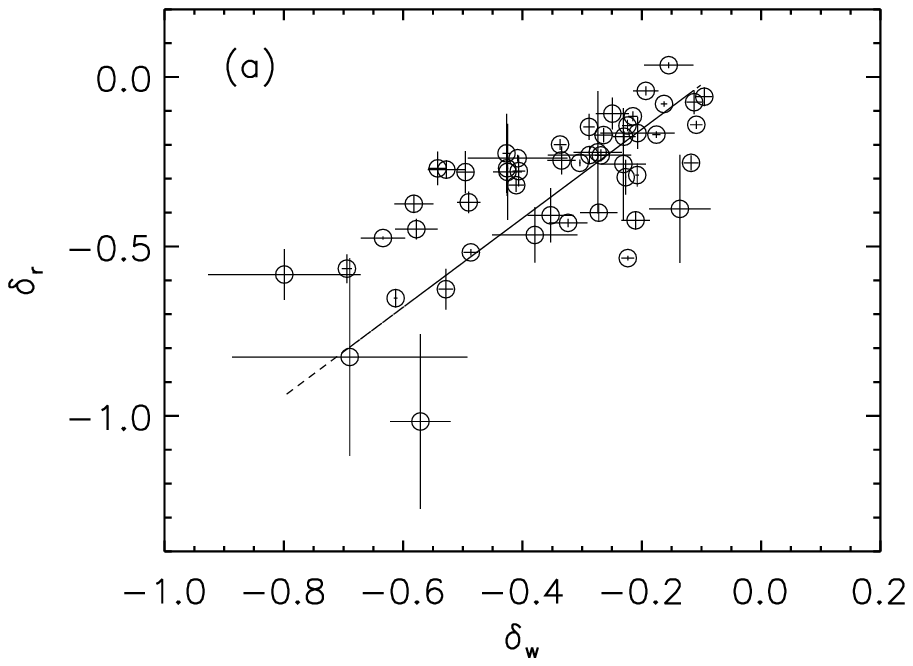}}
\resizebox{5.25cm}{4cm}{\includegraphics{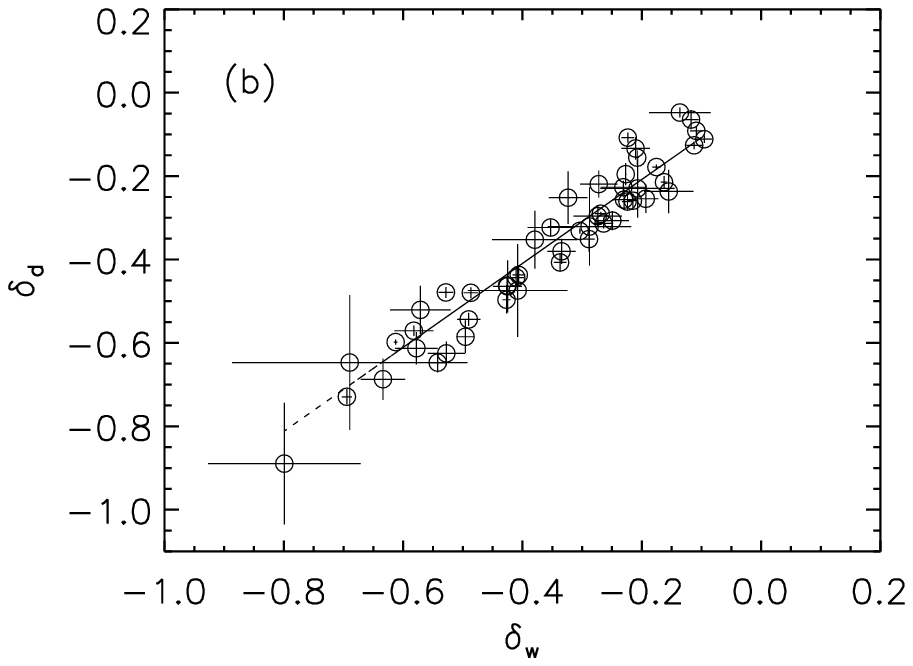}}
\resizebox{5.25cm}{4cm}{\includegraphics{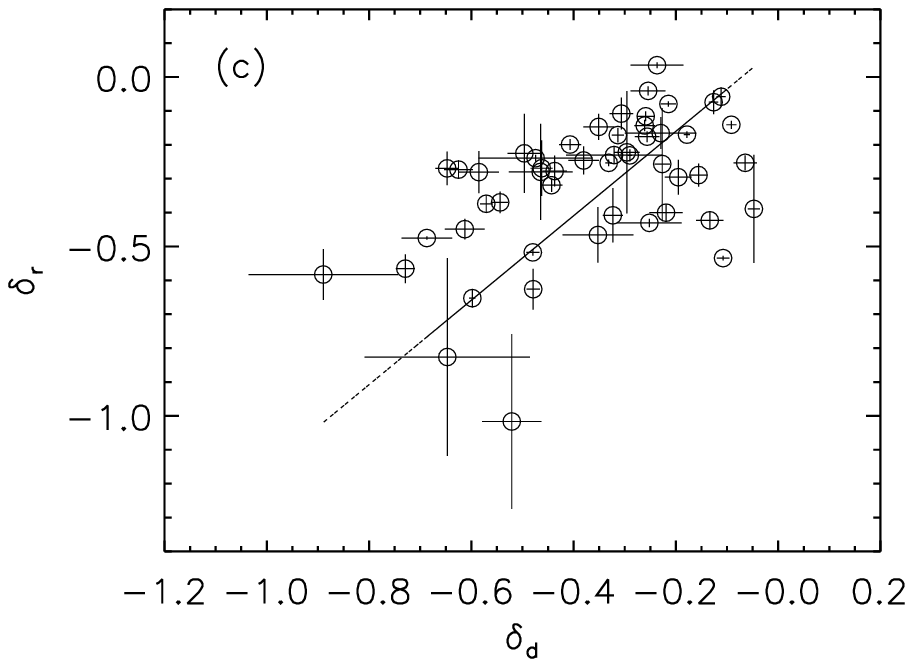}}
\caption{Relationships of the three power-law indices $\delta _{r}$ vs. $\delta _{w}$ (a), $\delta _{d}$ vs. $\delta _{w}$ (b), $\delta
_{r}$ vs. $\delta _{d}$ (c). The dashed lines are the regression lines.}
\label{}
\end{figure}

\begin{deluxetable}{ccccccccc}
\tabletypesize{\scriptsize}  \tablecaption{Correlations of the Three Power-law Indices versus Spectral Parameters
\label{tbl-4}}\tablewidth{0pt} \tablehead{\colhead{Parameter} & \colhead{$R_{S}$}  & \colhead{$P_{S}$}&\colhead{Parameter} &
\colhead{$R_{S}$} & \colhead{$P_{S}$}&\colhead{Parameter} & \colhead{$R_{S}$}  & \colhead{$P_{S}$}} \startdata

$\delta _{w}$-$\alpha$ & -0.57 &  $1.41\times 10^{-5}$ & $\delta _{w}$-$\beta$ & 0.39 &  $4.42\times 10^{-3}$&$\delta _{w}$-$E_{p}$ & 0.20 &  $1.05\times 10^{-1}$\\
$\delta _{r}$-$\alpha$ & -0.24 &  $8.99\times 10^{-2}$ & $\delta _{r}$-$\beta$ & 0.34 &  $1.37\times 10^{-2}$&$\delta _{r}$-$E_{p}$ & 0.19 &  $1.77\times 10^{-1}$\\
$\delta _{d}$-$\alpha$ & -0.62 &  $1.04\times 10^{-6}$ & $\delta _{d}$-$\beta$ & 0.32 &  $2.22\times 10^{-2}$&$\delta _{d}$-$E_{p}$ & 0.21 &  $1.44\times 10^{-1}$\\
\enddata
\tablecomments{$R_{S}$ and $P_{S}$ denote the Spearman rank-order correlation coefficient and significance, respectively.}
\end{deluxetable}

\subsection{THE RELATIONS AMONG THE POWER-LAW INDICES AND THE BAND SPECTRAL PARAMETERS}
Since the power-law indices reflect the spectral evolution of GRB
pulse we wonder whether they are related to the spectral parameters
or not. With the temporal and spectral parameters obtained in the
previous section, we can examine the relationships between the
power-law indices and the spectral shape parameters, low-energy
index $\alpha$, high-energy index $\beta$, and peak energy $E_{p}$.
Illustrated in Figure 5 are the scatter plots of the power-law
indices versus $\alpha$. The Spearman rank-order correlation
parameters are listed in Table 4. The regression analysis gives the
best-fitting lines: (1) $\delta _{w} = (- 0.63 \pm 0.01) + (0.40 \pm
0.01) \alpha$; (2) $\delta _{r} = (- 0.70 \pm 0.02) + (0.50 \pm
0.02) \alpha$; (3) $\delta _{d} = (- 0.62 \pm 0.01) + (0.39 \pm
0.01) \alpha$.

The strong anticorrelated relations are identified between $\delta
_{d}$ and $\alpha$ as well as $\delta _{w}$ and $\alpha$. However,
there seem no correlation between $\delta_{r}$ and $\alpha$  and the
dispersion is much larger than the other two parameter pairs. In
addition, the slope and intercept of $\delta _{r}$ and $\alpha$ has
apparent difference compared with that of $\delta _{d}$ and
$\alpha$. A possible interpretation of the phenomenon is that the
mechanism causing the dependence of the rise width on energy might
be different from the other temporal properties on energy. Another
possible reason is that the mechanism of producing the rise phase is
different from that of the decay phase.
%\begin{figure*}
% \vspace{202pt} %{plag13windexplot.ps}
% \begin{center}
%\resizebox{5.25cm}{4cm}{\includegraphics{Fig13.PS}}
%\resizebox{5.25cm}{4cm}{\includegraphics{Fig14.PS}}
%\resizebox{5.25cm}{4cm}{\includegraphics{Fig15.PS}}
%
%\end{center}
%\caption{Power-law indices $\delta _{w}$ (a), $\delta _{r}$ (b) and $\delta _{d}$ (c) vs. low-energy power-law index $\alpha$.  The dashed
%lines are the best fitting lines.}
%\end{figure*}

\begin{figure}
\centering
\resizebox{5.25cm}{4cm}{\includegraphics{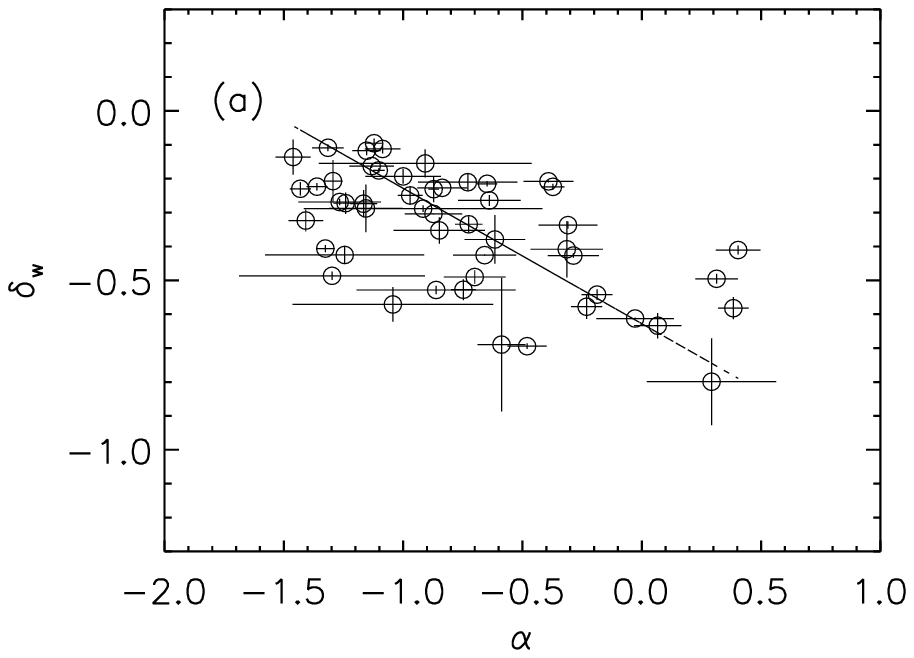}}
\resizebox{5.25cm}{4cm}{\includegraphics{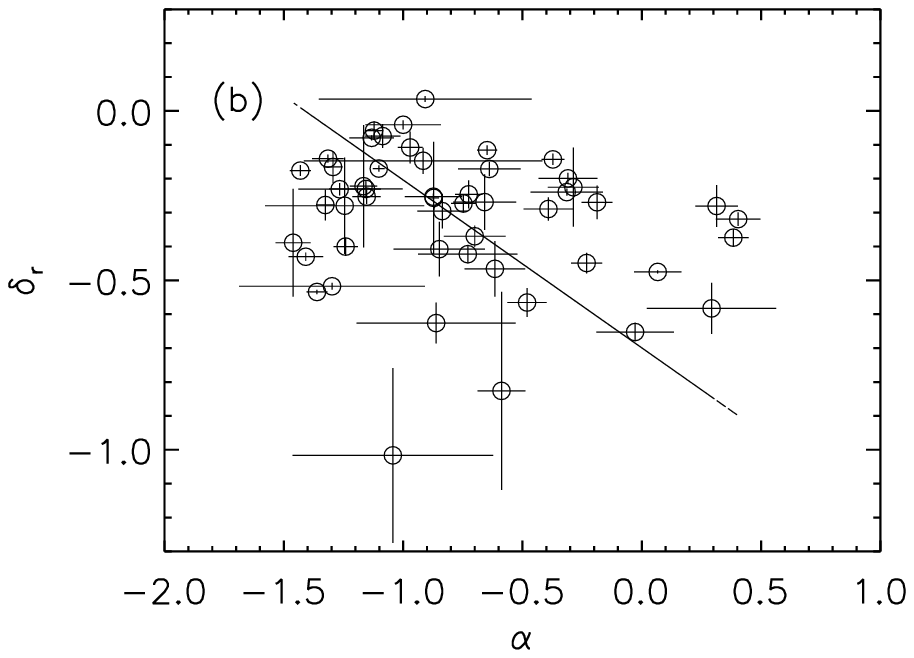}}
\resizebox{5.25cm}{4cm}{\includegraphics{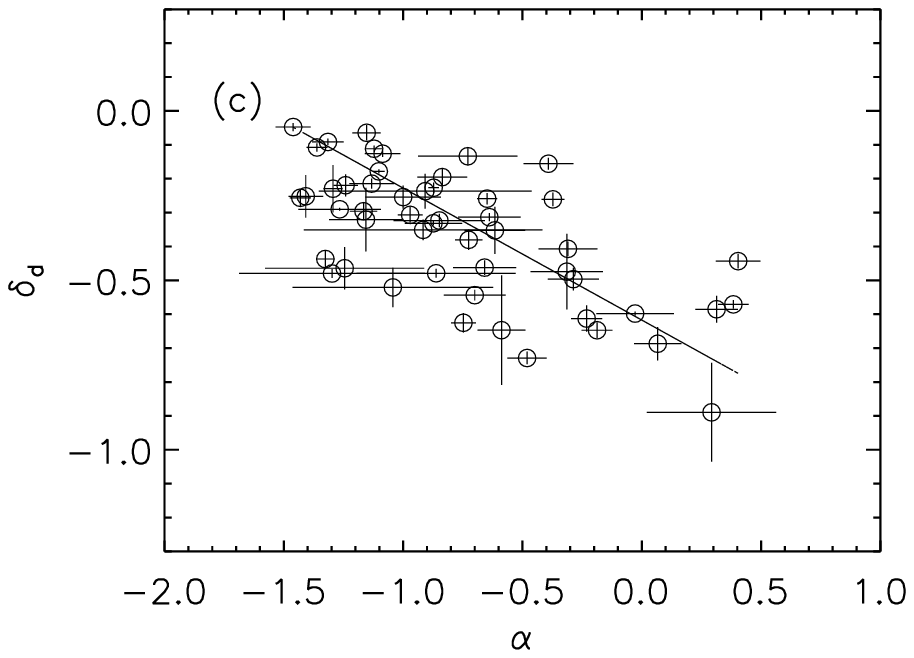}}
\caption{Power-law indices $\delta _{w}$ (a), $\delta _{r}$ (b) and $\delta _{d}$ (c) vs. low-energy power-law index $\alpha$.  The dashed
lines are the best fitting lines.}
\label{}
\end{figure}

%\begin{figure*}
% \vspace{202pt}
% \begin{center}
%\resizebox{5.25cm}{4cm}{\includegraphics{Fig16.PS}}
%\resizebox{5.25cm}{4cm}{\includegraphics{Fig17.PS}}
%\resizebox{5.25cm}{4cm}{\includegraphics{Fig18.PS}}
%
%\end{center}
%\caption{Power-law indices $\delta_{w}$ (a), $\delta _{r}$ (b), and $\delta _{d}$ (c) vs. high-energy power-law index $\beta$. The dashed
%lines are the best fitting lines.}
%\end{figure*}

\begin{figure}
\centering
\resizebox{5.25cm}{4cm}{\includegraphics{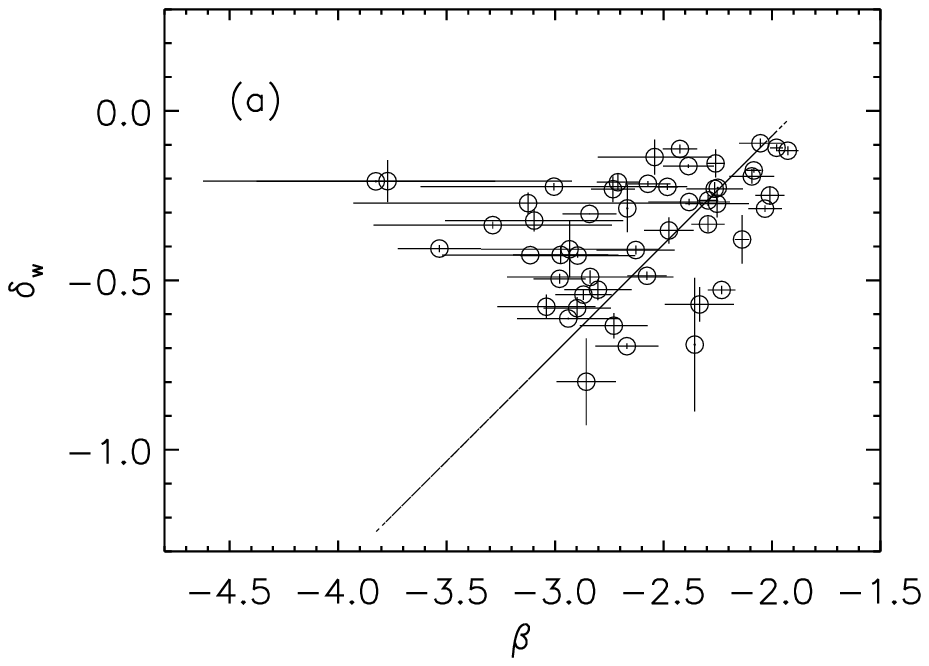}}
\resizebox{5.25cm}{4cm}{\includegraphics{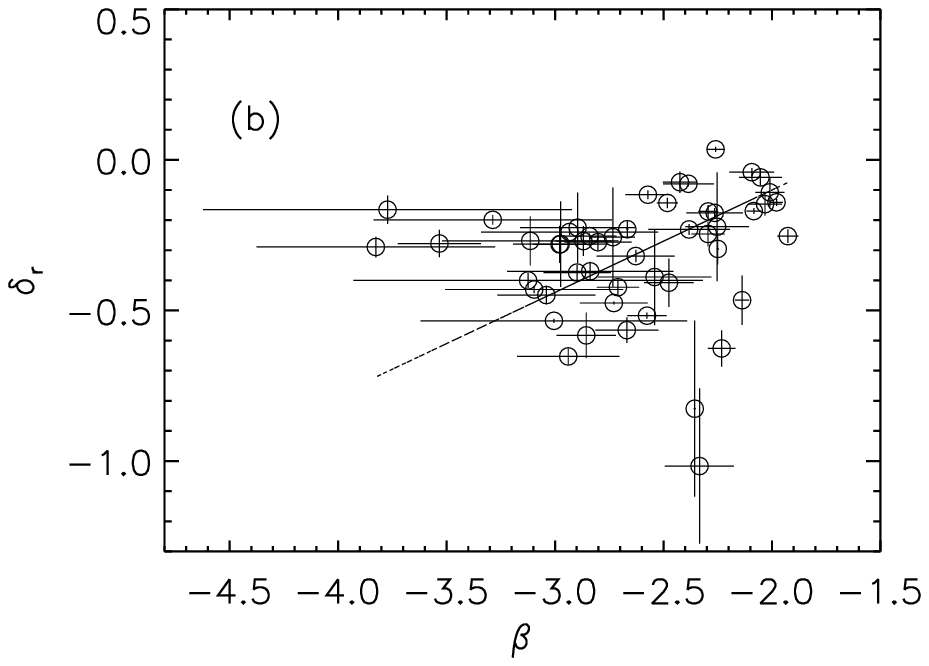}}
\resizebox{5.25cm}{4cm}{\includegraphics{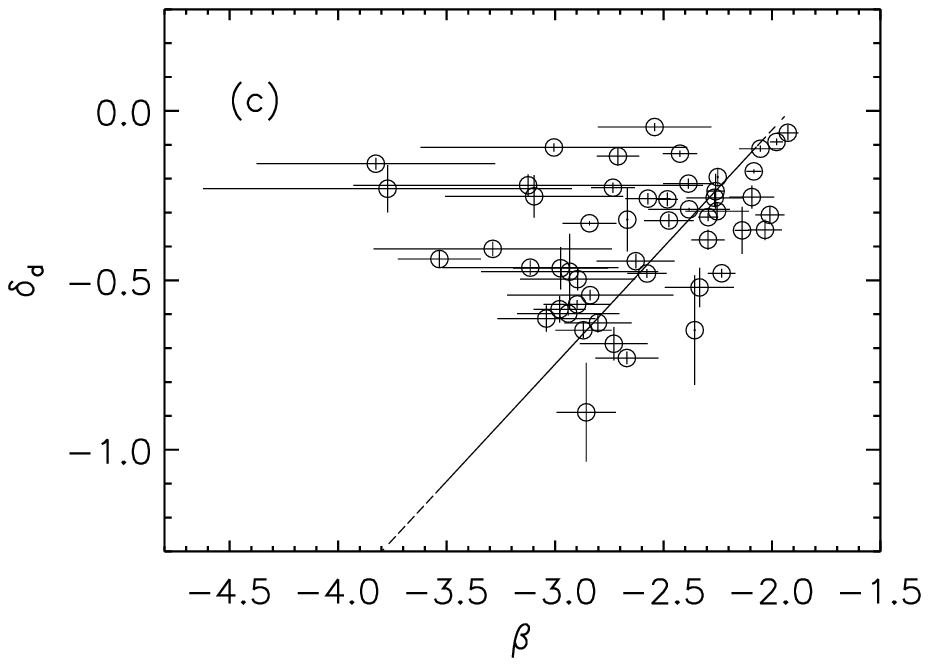}}
\caption{Power-law indices $\delta_{w}$ (a), $\delta _{r}$ (b), and $\delta _{d}$ (c) vs. high-energy power-law index $\beta$. The dashed
lines are the best fitting lines.}
\label{}
\end{figure}

%\begin{figure*}
% \vspace{202pt} %{plag13windexplot.ps}
% \begin{center}
%\resizebox{5.25cm}{4cm}{\includegraphics{Fig19.PS}}
%\resizebox{5.25cm}{4cm}{\includegraphics{Fig20.PS}}
%\resizebox{5.25cm}{4cm}{\includegraphics{Fig21.PS}}
%
%\end{center}
%\caption{Power-law indices $\delta _{w}$ (a), $\delta _{r}$ (b), and $\delta _{d}$ (c) vs. peak energy $E_{p}$. No correlation is
%apparent.}
%\end{figure*}

\begin{figure}
\centering
\resizebox{5.25cm}{4cm}{\includegraphics{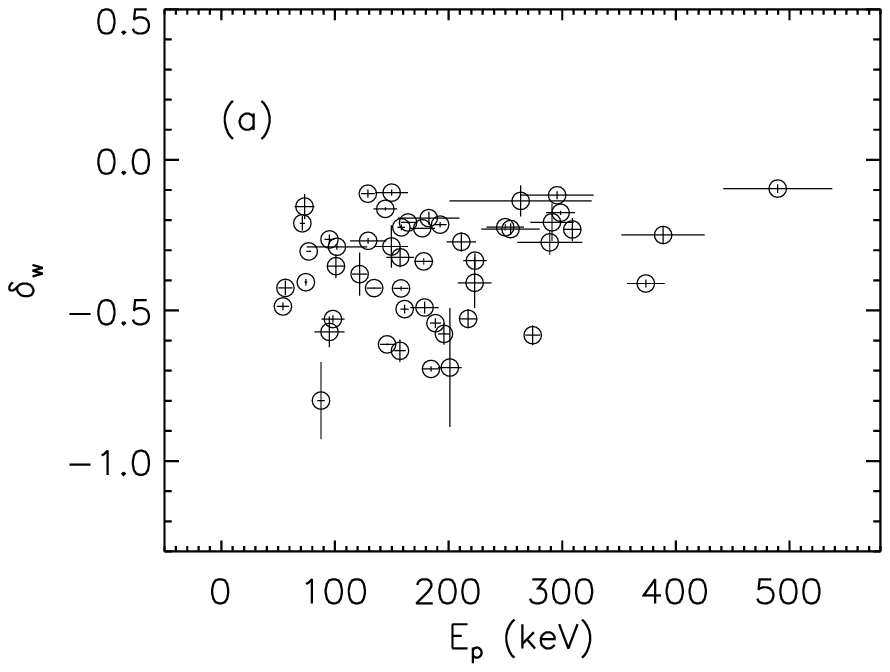}}
\resizebox{5.25cm}{4cm}{\includegraphics{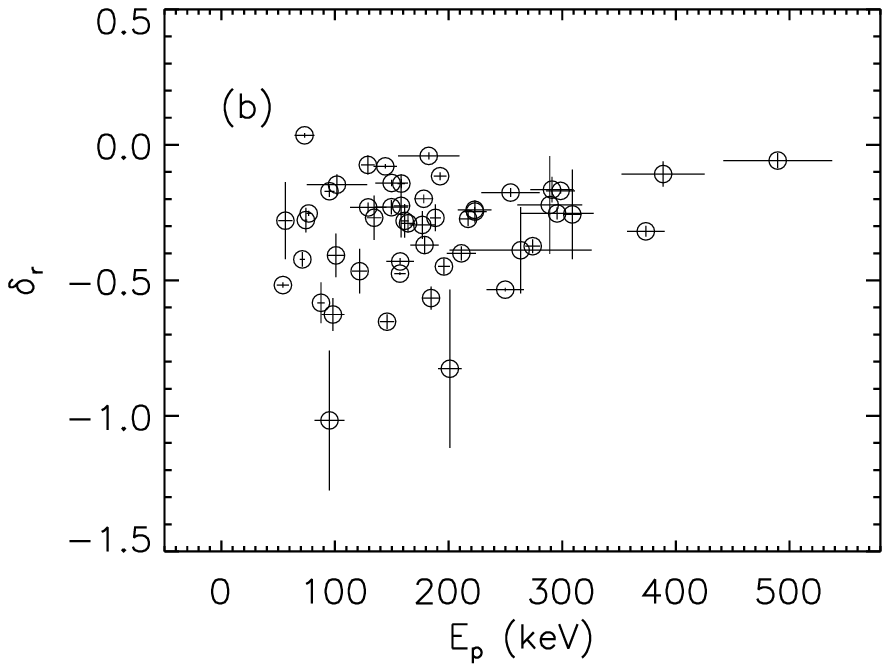}}
\resizebox{5.25cm}{4cm}{\includegraphics{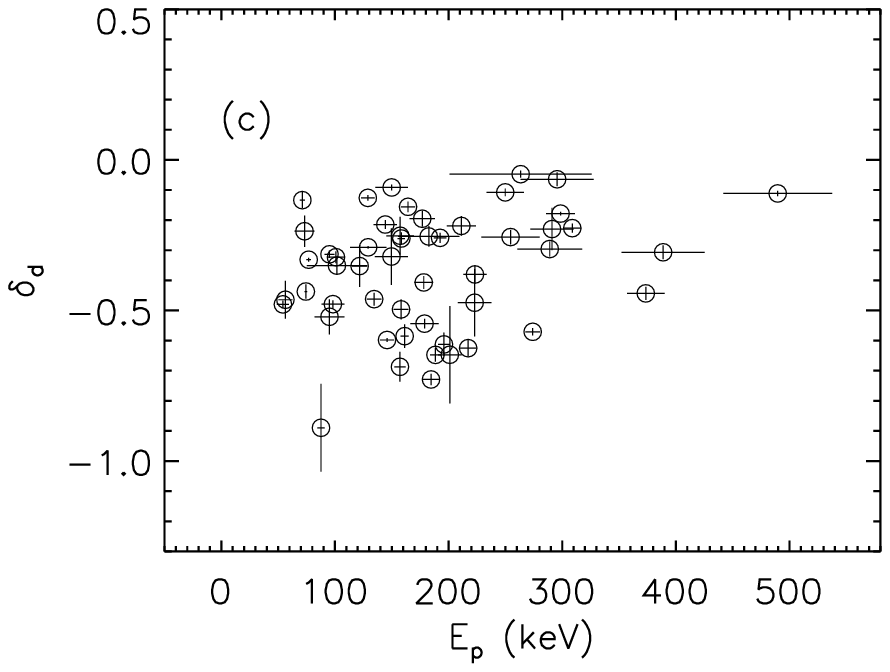}}
\caption{Power-law indices $\delta _{w}$ (a), $\delta _{r}$ (b), and $\delta _{d}$ (c) vs. peak energy $E_{p}$. No correlation is
apparent.}
\label{}
\end{figure}

Examining of the power-law indices versus the other two shaped
parameters $\beta$ and $E_{p}$, shows that all of the power-law
indices are weakly correlated with the $\beta$ (see, Figure 6 and
Table 4). However, no apparent correlations are found between the
three power-law indices and $E_{p}$ (see, Figure 7 and Table 4).

\section{IMPLICATIONS OF THE STATISTICAL CORRELATIONS}
The origin of the found correlations between the power-law indices and the Band-spectrum parameters could be related to the GRB pulse
formation mechanism. The mechanisms of the rise phase and decay phase of the pulse are different. The rise phase of the pulse is
believed to be related to gamma-ray active time of emission region, while the decay phase originates from the curvature effect. The
found strong $\delta_d-\alpha$ correlation and weak $\delta_r-\alpha$ correlation also suggest the different origins of
the rise phase and decay phase. It is interesting to speculate the physical origin of the strong $\delta_d-\alpha$ correlation. Since
$\delta_d$ is a quantity describing the decay phase, we consider the effect of the curvature effect on the energy dependence of the
pulse.

\subsection{The energy dependent of pulse width due to the curvature effect}
In order to study the energy dependent of pulse width due to the curvature effect, we first calculate the decay profile of a pulse
from a thin spherical surface expanding with relativistic speed. Define an emissivity
\bea
{j'_{E'} } = {\sum}^\prime_{E'} \delta (t' - {t_0'})\delta (R' -
{R_0'}),
 \eea
where ${\sum}^\prime_{E'}$ is the surface brightness per unit photon
energy interval in the comoving frame. Since the observed GRB
spectrum is broken power law form, it is not unreasonable to assume
${\sum}^\prime_{E'}$ to be the following form:
 \bea
{\sum}_{E'}^\prime  = {\sum}_0^\prime\left\{\begin{array}{ll}
({E'/E_p})^{1+\alpha} &  E'<E_p'\\
({E'/E_p})^{1+\beta} &  E'>E_p'\end{array} \right.,
  \eea
where ${\sum}_0^\prime$ is the normalized coefficient, $E_p'$ is the intrinsic cutoff energy (comoving frame), and $\alpha$ and $\beta$
are the low and high energy photon number spectrum slopes, respectively. As we know, more generally GRB spectrum is described
by the so-called Band function:
 \bea
{\sum}_{E'}^\prime  = {\sum}_0^\prime\left\{ {\matrix{
   {{{\left( {{{E'} \over {100keV}}} \right)}^{ 1+\alpha }}\exp \left[ { - {{E'(2 + \alpha )} \over {{E'_p}}}} \right]} & {E' < {E'_p}(\alpha  - \beta )/(2 + \alpha )}  \cr
   {{{\left[ {{{(\alpha  - \beta ){E'_p}} \over {(2 + \alpha )100keV}}} \right]}^{\alpha  - \beta }}\exp (\beta  - \alpha ){{\left( {{{E'} \over {100keV}}} \right)}^{ 1+\beta }}} & {E' > {E'_p}(\alpha  - \beta )/(2 + \alpha )}  \cr
 } } \right..
  \eea
However we will find below, the broken power law spectra gives more explicit physical meaning than the Band function.

The observed photon flux density can be given by
 \bea
{F_{E} } = {1\over 4\pi D_L^2}\int {{j_{E}}dV  = {1\over 2
D_L^2}\int\!\!\!\int
 {{j'_{E'}}{D^2}R^2dR d\mu } },
 \eea
Here $D=1/[\Gamma(1-\beta_\Gamma \mu)]$ is the Doppler factor and
the transformations $j_E=D^2j'_{E'}$ and $E=DE'$ are used. The
photon flux per unit photon energy can be given by
 \bea
N_E(T)={F_E \over E}.
 \eea

The observed time is $T=t-R\mu/c$ and $T=0$ was chosen as the time
of arrival at the observer of a photon emitted at $\mu=1$, $t=t_0$
and $R=R_0$. Thus we have $t_0=R_0/c$ and
$D\mid_{R=R_0}=2\Gamma/[1+(T/T_{ang})]$ where
$T_{ang}=R_0/(2\Gamma^2c)$. Note $\beta_\Gamma=1-1/2\Gamma^2$ is
used in the above derivation. Using $d\mu = cdt/R $,
$dt'=dt/\Gamma$, $dR'=\Gamma dR$, and
$\delta(t'-t_0')\delta(R'-R_0')=\delta(t-t_0)\delta(R-R_0)$, one can
get
 \bea
N_E(T)&=&{F_E \over E }\nonumber\\
&=&N_0D^2\left\{\begin{array}{ll} ({E'/E_p'})^{\alpha+1} & E'<E_p'\\
({E'/E_p'})^{\beta+1} &  E'>E_p'\end{array} \right.\\
&=&N_0D^2\left\{\begin{array}{ll} (\f{E}{DE_p'})^{\alpha+1} & E<DE_p'\\
(\f{E}{DE_p'})^{\beta+1} &  E>DE_p'\end{array} \right.,
 \eea
where $N_0=\Sigma'_{0}R_0c/(2D_L^2E)$ is the normalized coefficient. One can find, from the above equation, the observed photon flux
behaves as $N_E\propto(1+T/T_{ang})^{\alpha-1}$ or $N_E\propto(1+T/T_{ang})^{\beta-1}$ for a given observed photon
energy $E$.
%and $N_0=(1/100keV)\Sigma'_{0}R_0c/(2D_L^2)$ for Band function case,
%and the observed time $T$ is given by $T=R(1-\mu)/c$ and thus
%$\mu=1-Tc/R$.
%Replace $\mu$ with T, we can get the light
%curve.
The observed photon flux within an energy interval of $E_1-E_2$ can be written as
% \bea
%N(T)&=&\int^{E_2}_{E_1}{F_E \over E }dE %\\
%%&=&N_0\int^{E_2}_{E_1}D^2dE\left\{\begin{array}{ll}
%%({E'/E_p})^{-\alpha} &  E'<E_p'\\
%%({E'/E_p})^{-\beta} &  E'>E_p'\end{array} \right.,
% \eea
 \bea \label{eq:count rate}
N(T)&=&\int^{E_2}_{E_1}{F_E \over E }dE
 \eea

The decay phase of light curve in GRB is usually believed to be the
result of curvature effect, i.e., the photons at higher latitude
arrive at the observer more later with lower flux due to the
spherical emitting area and Doppler effect. If we consider the light
curve for a given observed energy, then the comoving (or intrinsic)
photon energy contributing to the given observed photon energy moves
toward higher energy end with the decay of light curve due to the
Doppler effect. The movement would experience a break so long as the
given energy is less than $E_p$ since the intrinsic spectrum is a
broken power law form. The light curve will first decay with
$N_E\propto (1+T/T_{ang})^{\alpha-1}$, and then with $N_E\propto
(1+T/T_{ang})^{\beta-1}$ so long as the given observed energy
satisfies $E<E_p$. The transition time, defined as $T_{ep}$, is the
time when the intrinsic photon energy contributing to the flux at
$E$ just moves to $E_p'$. However, usually the full width half
maximum (FWHM, defined as $T_{1/2}$) of a pulse or the width when
the flux decays to $1/e$ of peak flux (defined as $T_{1/e}$), as
used in this paper, are considered to be the widthes of the pulse.
These artificial definitions of pulse width could lead to the result
that the energy dependence of pulse width with energy does not be
\emph{measured} though it can present. The reason is as follows.
Following the the definition of $T_{1/2}$, it can be given by
$T_{1/2}=[2^{-1/(\alpha-1)}-1]T_{ang}$. Consider an observed photons
with energy $E<E_p$. The intrinsic contribution to it is initially
from photons of $E'_a=E_a/(2\Gamma)<E'_p$. When the photon energy
increase from $E'_a$ to $E'_p$ as the decay of Doppler factor in
high latitude region, the Doppler factor is
$D_{E_a}=E_a/E'_p=2\Gamma/(1+T_{ep}/T_{ang})$ and the time is
$T_{ep}$, so $T_{ep}$ is solved to be $T_{ep}=(2\Gamma
E'_p/E_a-1)T_{ang}$. If $T_{ep}>T_{1/2}$, i.e.,
$\alpha<1-ln2/ln(E_p/E_a)$, then the FWHM is only determined by the
low energy section, otherwise, the FWHM of light curve will be
affected by the high energy part of the intrinsic spectrum. If two
photons, say $E_a$ and $E_b$, both satisfy
$\alpha<1-ln2/ln[E_p/E_a(b)]$, then we can't \emph{measured} the
energy dependence of pulse width. Only if one of the photon energies
or both of them satisfy the condition $\alpha>1-ln2/ln[E_p/E_a(b)]$,
the energy dependence of pulse width would be \emph{measured}. If we
consider the time $T_{1/e}$, then $T_{ep}<T_{1/e}$ gives
 \bea \label{alpha}
\alpha>1-1/ln(E_p/E_a).
 \eea
In this condition, the closer to $E_p$ the photon energy (i.e., the
higher energy), the narrower the pulse width of this energy, shown
in Figure 8. This naturally explains the energy dependence of pulse
width. If two energy bands are both more energetic than $E_p$, then
the pulse widthes in these two bands will be the same, i.e., no
pulse broadening with energy presents in current consideration.

BATSE bursts, the typical $E_p$ is $\thicksim200$ keV, the
typical low energy slope is $-1$ and four energy channels are
(25-50, 50-100, 100-300, and 300-1000 keV) (Preece et al. 2000). The
condition (\ref{alpha}) give $E_p>121$ keV for the typical
parameters. Therefor for energy channels 1 and 2, the condition is
not satisfied and thus the pulse broadening in low energy band will
not be measured in the two channels. However, it is so only for
typical parameters. For a specific burst, $\alpha>1-1/ln(E_p/E_a)$
may be satisfied. Channel 3 crosses the typical $E_p$, while channel
4 is larger than it and thus only decays with $N_E\propto
(1+T/T_{ang})^{\beta-1}$. Thus the pulse width will generally
decrease from energy channel 4 to channel 2. In current
consideration, our results indicate no broadening measured in
channels 1 and 2 for the bursts with typical parameters. However,
there are two possibilities to lead to the broadening in channels 1
and 2.

One possibility is that the shell of internal shock is not
a spherical symmetry, which is proposed by some authors, e.g.,
Fenimore \& Sumner (1997) and Kocevski, Ryde \& Liang (2003).
Kocevski, Ryde \& Liang 2003 found that only $\sim40\%$ bursts are
consistent with the spherical curvature. Their results point to a
picture that the shell is a prolate geometry for a large number of
bursts. In this case, the Doppler factor is decreasing faster than
the spherical geometry so that the condition of (\ref{alpha}) is
easier to satisfy. Thus the prolate shell could lead to the
broadening in channels 1 and 2. Another one is that the intrinsic
spectrum can be evolving. If the intrinsic spectrum is evolving,
e.g., $E_p$ decreasing with time, then the condition (\ref{alpha})
can be satisfied so that the pulses in channels 1 and 2 also broaden
with energy. However, we do not know how the intrinsic spectrum
evolves and how fast it does. The curvature effect naturally gives
rise to an observed spectral evolution due to the Doppler effect,
which gives an $E_p$ decay with $T^{-1}$ for a spherical curvature.
If intrinsic spectral evolution gives a steeper observed one than
$T^{-1}$, then it will be masked and thus yield to $T^{-1}$ given by
curvature effect.

\begin{figure}
\includegraphics[width=\columnwidth]{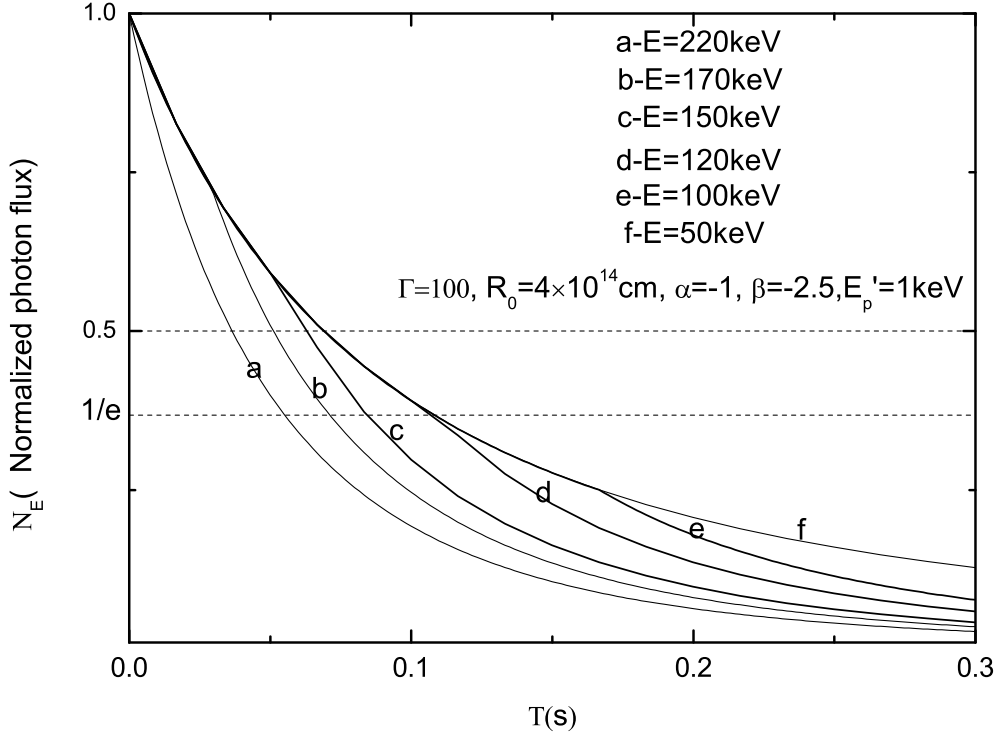}
 \caption{The light curves for different photon energy. Arrows mark the positions of $T_{ep}$.
The dashed lines label the positions of half and $1/e$ maximum flux.
Line a decays as $(1+T/T_{ang})^{\beta-1}$, lines b, c, d and e
first as $(1+T/T_{ang})^{\alpha-1}$ and then as
$(1+T/T_{ang})^{\beta-1}$ and line f as $(1+T/T_{ang})^{\alpha-1}$.
Lines b and c show energy dependence of width of decay while the
lines d and e do not due $T_{1/2}<T_{ep}$ if considering FWHM as the
decay width. } \label{fig:lcc}
\end{figure}

The above analysis assumes the intrinsic spectrum is broken power
law, so the light curve broken of decay phase at $T_{ep}$ can be
seen clearly (see Fig. \ref{fig:lcc}). Actually the GRB spectrum is
generally smoothly connected at the broken energy, namely, the Band
spectrum, and the observed light curve is usually from an energy
interval, not a single energy, so the observed decay phase is
usually smooth. Note that total pulse width in GRB is dominated by
decay phase. Thus the curvature effect+Band spectrum we consider
here can provide a natural explanation to the energy dependence of
GRB total pulse width, not only the decay width.

\subsection{$\delta_d-\alpha$ relation}
We have shown the Band spectrum+curvature effect is a reason of
pulse broadening with photon energy. Different Band function
parameters will cause the differences of the degree of pulse
broadening. However, the effects of the three Band function
parameters, $\alpha$, $\beta$ and $E_p$, on the degree of pulse
broadening are different. The main effect seems to come from the
low-energy slope. The reason is the low-energy section is usually
flatter than the high one, i.e., $\alpha>\beta$. Thus the decay,
$N_E\propto (1+T/T_{ang})^{\alpha-1}$, is also more shallow than
$N_E\propto (1+T/T_{ang})^{\beta-1}$ with the pulse decay width
dominated by $\alpha$, unless the given energy $E>E_p$. One can come
to the conclusion that the larger $\alpha$, the more significant the
broadening with energy. Suppose two extreme cases:
$\alpha=\beta$, i.e., the spectrum is a single power law; $\alpha=1$
and $\beta=\infty$, i.e., the spectrum is composed of a horizon line
and the vertical line. The pulse width would be the same for
different energy bands in the former case, while the broadening of
pulse with energy is the most significant in the latter case. Other
cases lies between the two cases. It is difficult to find directly
whether this is consistent with the observed anticorrelation in the
earlier sections. We thus make a calculation to test that.

In order to compare with the observations, we calculate the count
rate within four energy channels (25-50, 50-100, 100-300, and
300-1000 keV) with equation \ref{eq:count rate}. Further we
calculate the decay width of pulse within the four energy channels,
fit linearly the width-energy relation and then find the power law
slope, $\delta_d$. Figure 9 shows our theoretically calculated
$\alpha-\delta_d$ relation. Obviously, $\alpha$ and $\delta_d$ are
roughly linearly anti-correlated, which is consistent with the
statistical results of the observations. The fitted slope in Figure
9 is ~0.3, also roughly consistent with the statistical results.
This suggests that the curvature effect+Band spectrum could lead to
the observed $\delta_d-\alpha$ correlation. Strictly speaking,
$\delta_d$ and $\alpha$ are not linearly correlated for the
theoretical results (still see Figure 9), not completely consistent
with the observations. However, in our calculations, some plausible
values of the parameters, for instance the Lorentz factor and the
peak energy, are used, disregarding the variety of the parameter
value for different bursts, which could lead to the inconsistency.
Further whether the relation in physics is linearly anti-correlated
or other forms is not important. The observed relation results from
the sum of intrinsic physical factors, such as the Band spectrum,
curvature effect and some artificial factors , such as the energy
channel selection of BATSE and the pulse width definition, which is
difficult to figure out one by one. Here we only focus on the Band
spectrum and curvature effect. What we concern here is the effect of
the combination of them on the pulse. We find that the curvature
effect+Band spectrum can indeed result in such an anti-correlated
trend of $\alpha-\delta_d$.

\begin{figure}
\includegraphics[width=\columnwidth]{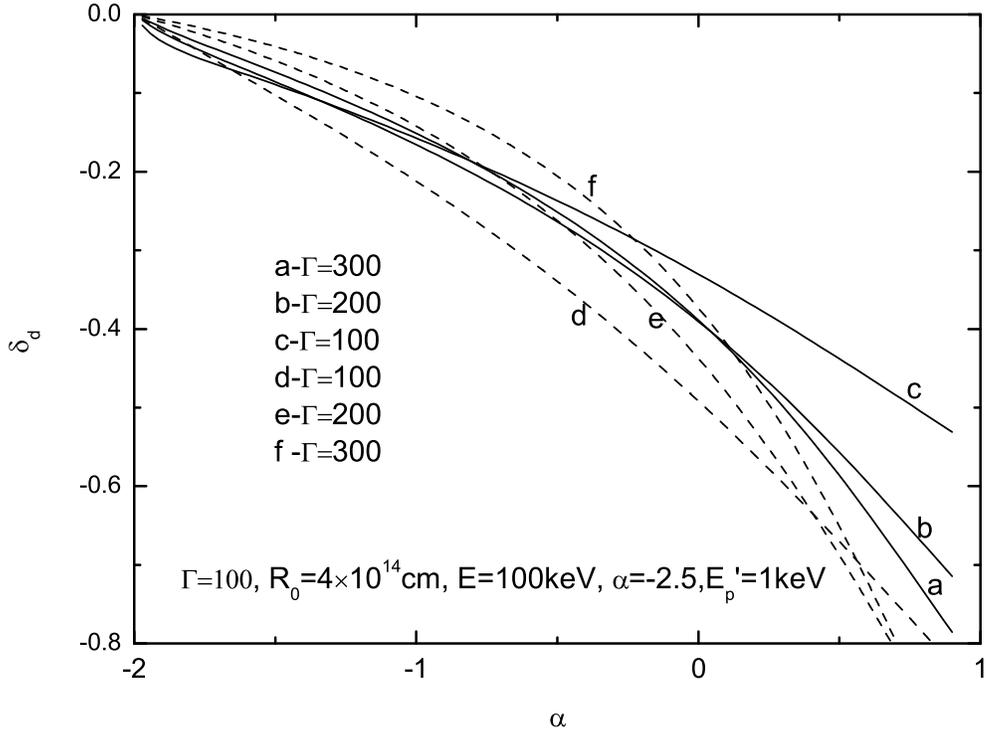}
\caption{$\delta_d-\alpha$ relation. The solid lines are for 4 channels of BATSE considered, while the dashed lines for 1, 2 and 3 channels case.
The latter is considered due to the fact that the fitted $\delta_r$,
$\delta_d$ and $\delta_w$ in statistical part of the paper are
obtained mainly using channels 1, 2 and 3.} \label{a-d relation}
\end{figure}

\section{Discussion and Conclusions}
In this paper, we have selected 51 well-separated long-duration FRED
GRB pulses from the available sample of BATSE and analyzed the pulse
temporal and spectral parameters with the aim to search for their
connections. Our analysis first shows that the pulse width, pulse
rise width and pulse decay width with energy all scale as a power
law with the power-law slopes of $\delta_w$, $\delta_r$ and
$\delta_d$, respectively. In addition, the slopes are correlated to
each other but the correlation between $\delta_w$ and $\delta_d$ is
much stronger than that of $\delta_w$ and $\delta_r$ as well as
$\delta_r$ and $\delta_d$, which may be due to the pulse width is
dominated by the decay phase and the rise and decay phase have
different origins.

Then we investigate the relations between the power-law indices and
the three Band spectral parameters. Our results show that the
power-law indices, $\delta_w$, $\delta_d$ are strongly correlated
with the low-energy index $\alpha$ and much less correlated with
high energy slope $\beta$. The $\delta_r$ weakly correlates with
$\beta$ but does not show apparent correlated with $\alpha$.

We do not find that there are apparent correlated relations between
the peak energy $E_p$ and three power-law indices based on our
current sample (see, Figure 7 and Table 4). However, we think the
demonstrations that $\delta_w$, $\delta_r$, and $\delta_d$ are
independent of $E_p$ may be weak only based on our current sample.
Several reasons may cause the case. Firstly, it is well known that
the BATSE $E_p$ distribution is narrow (e.g. Band et al. 1993,
Mallozzi et al. 1995, Schafer 2003), and thus it can be hard to
clearly identify the correlations in data drawn from a narrow
distribution. Secondly, the sample is likely biased toward low
$E_p$-valued pulses. The sampled pulses have been selected on the
basis of long durations and bright peak fluxes. Because of pulse
property correlations, such pulses tend to be softer than the mean
(Hakkila and Preece, 2011). This pulse selection bias can have
contributed to the narrowness of the sample. Lastly, some of the
highest $E_p$ pulses in this sample may have been incorrectly
identified as single pulses, when in actuality they represent merged
pulses. If the hard-to-soft pulse evolution demonstrated by Peng et
al. (2009a) is normal, then a pulse made from two merged pulsed will
experience a re-hardening when the second pulse kicks in (Hakkila
and Preece 2011; Ukwatta 2011), which can give these pulses higher
$E_p$ values than the other single pulses to which they are being
compared. Due to the bad fits to those merged pulses with given pulse
model we are not sure how the merging of pulses affects the $\delta$
values from the data analysis. But we can give some reasonable deductions.
Firstly, it seems that the smallest $\delta$ value from a merged
pulse occurs when pulses overlap completely (i.e. when they peak at
the same time). Otherwise, the pulses should appear to be longer in
all energy channels, and the pulse separation will make a larger
contribution to the pulse broadening than the energy-dependent
broadening embedded within it. Secondly, we find that the correlations between $E_p$ and
$\delta$s are still positive and the correlations get much weaker after removing those merged pulses with high $E_p$ values.
Therefore, if the merged pulses can cause high $E_p$ and the
positive correlations between $E_p$ and $\delta$s indeed exist the
overlapping pulses may also make the $\delta$s small. Based on the above analysis
we think the merged pulses would make the $\delta$ values decrease. In order to identify further the correlations
between $E_p$ and $\delta$s more wider energy band and cleaner singe pulse data are needed.

We further investigate the implications of the statistical
relations. We find that the energy dependence of decay width (also
the pulse width since it is dominated by the decay section) can be
caused by the curvature effect+Band spectrum. Usually we study the
observed light curve for a given energy or energy internal (e.g.,
BATSE has 4 energy channels). The light curve decays due to the
delay arrival of high-latitude photons with lower flux, i.e.,
curvature effect. For a given observed energy less than the peak
energy of Band spectrum, the intrinsic photon energy contributing to
the energy will move from low energy part of Band spectrum to high
energy one due to the lower Doppler factor at high latitude. This
will naturally lead to the energy dependence of decay width and thus
the pulse width. This in turn supports the decay phase is indeed due
to the curvature effect, or at least related to it. In addition,
different Band function parameters may lead to the different degree
of pulse broadening. The main effect seems to be from the low-energy
slope as analyzed above. The strong correlation between $\delta_d$
and $\alpha$ also appears to result from the curvature effect and
Band spectrum (see Fig.\ref{a-d relation}).

The rise phase of the GRB pulse contributes less to the pulse width
than the decay phase. However it is important and is considered to
be relevant to the hydrodynamic time of shock crossing shell.
Statistical studies have found that the rise phase is also related
to the time-resolved Band spectrum in single pulse. For instance,
some authors found that the time-resolved peak energy of pulse is
tracking the pulse profile within a pulse for some bursts (Kaneko et
al. 2006; Peng et al. 2009b, however, see Hakkila \& Preece 2011 for
different view). In current investigations, we didn't find
correlations between broadening of the rise phase and the
(time-integrated) Band spectrum parameters for different pulses,
suggesting that the formation of the rise phase is related to
emission mechanism, while its energy dependence is independent of
emission mechanism. The energy dependence of rise phase may arise
from the intrinsic spectral evolution. As pointed by Fenimore \&
Sumner (1997), the intrinsic spectral evolution affects the rise
phase much more significant than the decay phase.

A deduction of the correlation between $\delta_d$ and $\alpha$ we
found is that the energy dependence of pulse with single power law
(corresponding to $\alpha=\beta$ in Band spectrum) spectrum is
systematically weaker than that with Band spectrum. If it is
verified, this will support that the intrinsic Band spectrum and the
curvature effect is indeed an important factor leading to the photon
energy dependence of GRB pulses. Also it in turn supports that the
decay phase of pulse is produced by the curvature effect. We will
consider it in a future paper with Fermi/GBM or Konus-Wind data.

\section{Acknowledgments}
We thank the anonymous referee for constructive suggestions. This work was supported by the Natural Science Fund of Yunnan Province (2009ZC060M), the Key Program for Science Fund of the Education Department
of Yunnan Province (2011Z004), the Open Research Program of Key Laboratory for the Structure and Evolution of Celestial Objects (OP201106),
and the National Natural Science Foundation of China (no. 10778726).

\clearpage
\end{document}